\documentclass[aps,prl,reprint,superscriptaddress,showpacs]{revtex4-1}

\usepackage{graphicx}%
\usepackage{dcolumn}% Align table columns on decimal point
\usepackage{bm}% bold math
\usepackage{amsmath,amssymb}
\usepackage{color}
\begin{document}
%\preprint{kkk}

\title{Anomalous Quasiparticle Reflection from the Surface of a $^3$He-$^4$He Dilute Solution}

%\textcolor{magenta}{
% repeat the \author .. \affiliation  etc. as neededS

\author{Hiroki Ikegami}\email{hikegami@riken.jp}
\affiliation{The Center for Emergent Matter Science, RIKEN, Wako, Saitama 351-0198, Japan}

\author{Kitak Kim}
\affiliation{Department of Physics, KAIST,  Daejeon 34141, Republic of Korea}

\author{Daisuke Sato}
\affiliation{The Center for Emergent Matter Science, RIKEN, Wako, Saitama 351-0198, Japan}

\author{Kimitoshi Kono}
\affiliation{The Center for Emergent Matter Science, RIKEN, Wako, Saitama 351-0198, Japan}

\author{Hyoungsoon Choi}\email{h.choi@kaist.ac.kr}
\affiliation{Department of Physics, KAIST,  Daejeon 34141, Republic of Korea}

\author{Yuriy P. Monarkha}
\affiliation{Institute for Low Temperature Physics and Engineering, 47 Nauky Avenue, Kharkov 61103, Ukraine}

\date{\today}

\begin{abstract}
A free surface of a dilute $^3$He-$^4$He liquid mixture is a unique system where two Fermi liquids with distinct dimensions coexist: a three-dimensional (3D) $^3$He Fermi liquid in bulk and a two-dimensional (2D) $^3$He Fermi liquid at the surface.
To investigate a novel effect generated  by the interaction between the two Fermi liquids, mobility of a Wigner crystal of electrons formed on the free surface of the mixture is studied.
An anomalous enhancement of the mobility, compared with the case where the 3D and 2D systems do not interact with each other, is observed.
The enhancement is explained by non-trivial reflection of 3D quasiparticles from the surface covered with the 2D $^3$He system.
\end{abstract}

\pacs{}
%\keywords{}

\maketitle

Dimensionality is one of the defining characteristics that govern physical properties of systems.
Although a lot of experimental and theoretical investigations have clarified static and dynamic properties of quantum many-body problems, the studies so far have been mostly limited to the cases with a well-defined single dimension.
Mixed-dimensional systems, in which two quantum many-body systems with distinct dimensions coexist, are expected to exhibit nontrivial dynamics that never occur in either subsystem by itself.
Such systems are found in various fields, for example, electrons in noble metals \cite{book_Zangwill} and unconventional materials \cite{Plumb2014,Wang2012} (electrons inside and on the surface), and ultracold atoms \cite{Nishida2008,Iskin2010,Okamoto2017}, but in many cases, the two subsystems are not well understood or well distinguished.

The free surface of a dilute $^3$He-$^4$He mixture \cite{Dobbs00} is unique in this regard since well-characterized two-dimensional (2D) and three-dimensional (3D) fermionic systems of identical particles, i.e., $^3$He, coexist \cite{Guo71,Edwards75}.
When a $^3$He quasiparticle (QP) in the 3D system approaches the 2D $^3$He layer formed at the free surface and interacts with it, the interaction is expected to affect the reflection process and surface dynamics.
In this letter, we demonstrate that novel QP reflection from the surface manifests itself as anomalous enhancement of the mobility of a Wigner crystal (WC) of electrons trapped on the surface of the mixture.

\begin{figure}
\begin{center}
\includegraphics[width=0.9\linewidth,keepaspectratio]{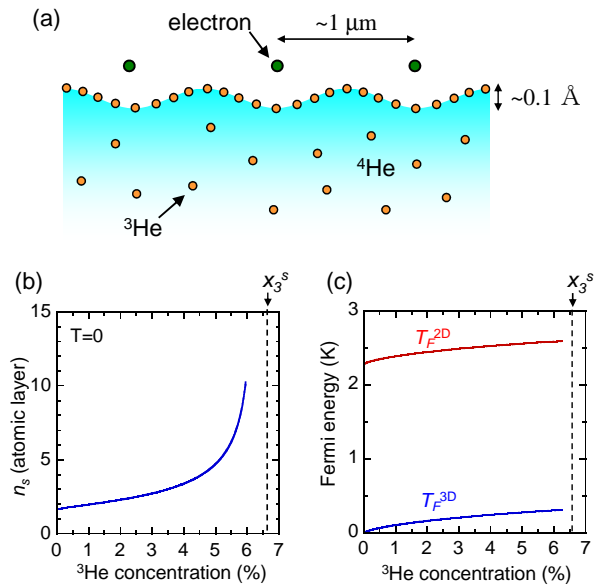}
\end{center}
\caption{\label{WC_and_surface_layer} (a) $^3$He-$^4$He mixture and WC formed on a free surface. A 2D $^3$He layer is formed at the free surface of the mixture as a result of the larger zero-point motion of $^3$He than that of $^4$He \cite{Andreev1966,Saam1971}. Electrons are trapped about 10 nm above the surface. At low temperatures, the electrons form a WC dressed with a DL. The lattice constant of the WC is $a=$ 0.93 $\mu$m for our electron density of $n_e \sim$ 1.35$\times$10$^{12}$ m$^{-2}$, while the depth of the DL is only $\delta \sim$ 0.1 \AA .
(b) Areal density of the 2D $^3$He layer $n_s$ (in the unit of atomic layers with monolayer density 6.4$\times$10$^{18}$ m$^{-2}$) as a function of $x_3$ at $T=0$. This graph is based on the analysis by  Guo {\it et al.} of their surface tension data \cite{Guo71} (see Supplementary Material for details \cite{suppl}). The surface is covered by a monolayer of $^3$He even at a very small $x_3$ ($\sim$ 200 ppb). $n_s$ diverges approaching the saturation concentration ($x_3^s=$ 6.7\%).
(c) $T_F^{\rm 2D}$ and $T_F^{\rm 3D}$ as a function of $x_3$. (see Supplementary Material \cite{suppl}).
}
\end{figure}

We consider the mixture with a free surface.
The 3D $^3$He system is formed in bulk; $^3$He atoms, which are soluble in liquid $^4$He at concentrations of $^3$He $x_3$ up to $\sim$ 6.7\% \cite{Yorozu1992}, behave as a dilute 3D Fermi liquid in the background superfluid $^4$He  \cite{Dobbs00}.
At the surface, $^3$He atoms are bound to the surface to form a dense 2D layer [Fig. \ref{WC_and_surface_layer}(a)] \cite{Andreev1966,Saam1971}, showing the 2D Fermi liquid behavior \cite{Guo71,Edwards75}.
With increasing $x_3$, the thickness of the 2D $^3$He layer increases in the range of several atomic layers except at $x_3 \sim$ 0 and $\sim$ 6.7\% [Fig. \ref{WC_and_surface_layer}(b)] \cite{Guo71,Edwards75,Review_ProgLowTempPhys}, while the mean atomic distance is comparable to that of bulk pure $^3$He and is almost unchanged with $x_3$.
Therefore, the Fermi temperature $T_F^{\rm 2D}$ is high ($\sim$ 2 K) without significant variation with $x_3$ [Fig. \ref{WC_and_surface_layer}(c)].
This is in contrast to the low Fermi temperature of the 3D $^3$He $T_F^{\rm 3D}$ ($<$ 0.4 K) owing to the lower density in the mixture.
Just as the 2D and 3D $^3$He systems have been investigated as the prototypical Landau's Fermi liquid in each dimension \cite{Dobbs00,Guo71,Edwards75} owing to their cleanliness, the mixed-dimensional system presented in this work will serve as an ideal model system to study the interplay between the two subsystems.

Electrons on a free surface of liquid helium \cite{Book2DE,BookMonarkha} are often used as a sensitive microscopic probe for investigating fundamental properties of elementary excitations in quantum fluids.
The electrons undergo a transition to a WC at a certain low temperature, where the crystallization generates a commensurate deformation of the helium surface called a dimple lattice (DL) [Fig. \ref{WC_and_surface_layer}(a)] \cite{Book2DE,BookMonarkha}.
The emergence of the DL strengthens the coupling of the WC to the liquid, making the transport of the WC sensitive to properties of elementary excitations in the liquid \cite{Deville1988,Shirahama97,Ikegami06,Monarkha1997}.
This feature has been utilized to reveal, particularly, the specular nature of QP reflection from the surface in normal and superfluid $^3$He~\cite{Shirahama97,Ikegami06,Monarkha1997}.
In this study, we similarly measure the mobility of a WC over the $^3$He-$^4$He liquid mixture down to $\sim$ 10 mK over a wide range of $x_3$ (up to 6.1\%) to investigate the nature of QP dynamics near the surface.
(So far there has been an experimental study of WC mobility only at $x_3=$ 0.5\% \cite{Yayama_JLTP2014}).

The mobility is measured with the Sommer$-$Tanner technique \cite{Sommer71} in the Corbino geometry at a vertical magnetic field $B$ of 380--1100 Gauss.
The Corbino disk consists of two concentric electrodes 18.0 and 11.9 mm in diameter and is attached to the ceiling of the sample cell.
A bottom circular electrode, which is located 3.0 mm below the Corbino disk, is used to provide a vertical pressing electric field $E_{ \bot }$.
The free surface is set at a midway between the Corbino disk and the bottom electrode, and electrons are deposited on it.
The longitudinal conductivity $\sigma_{xx}$ is measured by applying an ac voltage $V_{\rm ac}$ of frequency $f=$ 214 kHz to the inner electrode of the Corbino disk and recording the induced current $I_{\rm out}$ on the outer electrode.
A small $V_{\rm ac}$($=$ 1 mV$_{\rm rms}$) is used to avoid nonlinear effects.
The mobility $\mu$ is deduced from the Drude relation $\sigma_{xx}=en_e \mu/[1+(\mu B)^2]$, where $n_e$ is the electron density.
Mobility at each $x_3$ is measured at a certain value of $n_e$ and $E_{ \bot }$ in the range of $n_e=$ (1.33--1.40)$\times$10$^{12}$ m$^{-2}$ and $E_{ \bot }=$ (2.02--2.08)$\times$10$^{4}$ V/m.
We use a rather high $E_{ \bot }$ to avoid the WC from decoupling from the DL easily by the ac drive.
The magnitudes of the mobility are calibrated by multiplying by a factor of about unity (0.94--1.14) so that the mobility agrees with the theoretical mobility of highly correlated electrons in the ripplon scattering regime \cite{BookMonarkha} above $T_m$ (see Supplementary Material \cite{suppl}).
The mixture is cooled to $\sim$ 10 mK with a heat exchanger made of packed silver powder.
The atomic concentration of $^3$He $x_3$ is determined from the amounts of $^3$He and $^4$He introduced into the cell.
We also monitor $x_3$ via the dielectric constant of the mixture using a parallel plate capacitor immersed in the liquid.

\begin{figure}
\begin{center}
\includegraphics[width=0.7\linewidth,keepaspectratio]{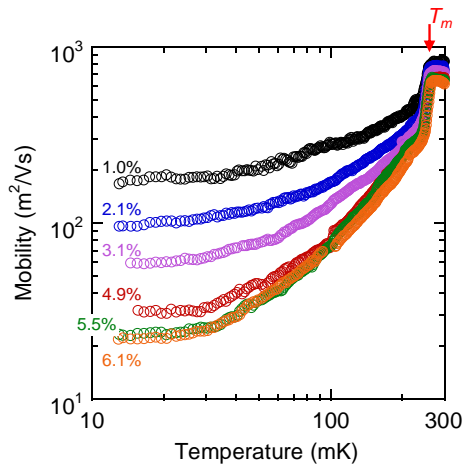}
\end{center}
\caption{\label{temp_dep_mobility}  Mobility of the WC as a function of $T$ for different $^3$He concentrations. $T_m$ indicates the transition temperature to the WC phase.
}
\end{figure}

Figure \ref{temp_dep_mobility} shows the mobility $\mu$ as a function of temperature $T$ at different $x_3$.
With decreasing $T$, $\mu$ exhibits a sudden drop at the transition temperature to the WC, $T_m \sim$ 260 mK, due to the formation of a DL, followed by a further reduction at lower $T$.
In the WC phase, two features are found:
the mobility at each $x_3$ asymptotically approaches a constant value below several tens of mK, and $\mu$ is significantly suppressed when $x_3$ is increased.

\begin{figure}
\begin{center}
\includegraphics[width=0.95\linewidth,keepaspectratio]{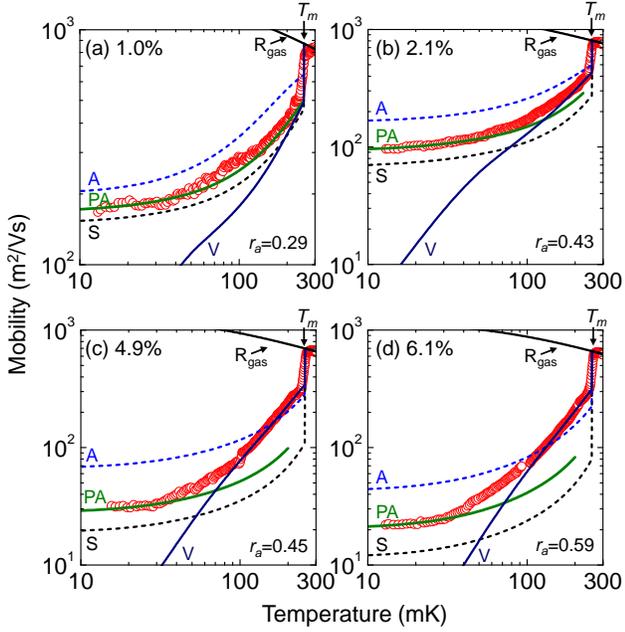}
\end{center}
\caption{\label{mu_T_with_theory} Experimental mobility of the WC compared with the theoretical calculations.  The experimental data are the same as those shown in Fig. \ref{temp_dep_mobility}. Theoretical mobilities limited by the viscosity (V), specular reflection (S), accommodation process (A), partial accommodation process (PA), and ripplon scattering in the electron gas regime (R$_{\rm gas}$) are shown for $n_e=$ 1.35$\times$10$^{12}$ m$^{-2}$ and $E_{ \bot }=$ 2.05$\times$10$^{4}$ V/m. (a) $x_3=$ 1.0, (b) 2.1, (c) 4.9, and (d) 6.1\%.
An error in the theoretical curves associated with the deviation of $n_e$ and $E_{\bot}$ used in the calculation from the actual values is estimated to be 3\% at most.
At $x_3=$ 1.0\%, the system is in the ballistic regime at temperatures just below $T_m$ because of the long mean free path, and therefore the viscous regime is not observed.
}
\end{figure}

Right below $T_m$, $\mu$ is limited by the viscosity of the bulk mixture $\eta$.
In this regime, $\mu$ is determined by the viscous drag force acting on the DL moving together with the WC \cite{Monarkha1997,Monarkha2012,BookMonarkha}.
We evaluate the theoretical mobility at our measurement frequency (214 kHz) using experimentally known values of viscosity $\eta$ \cite{Kuenhold1972,VanDerBoog1981,Konig1994}, density $\rho$ \cite{Edwards1969}, and surface tension $\alpha$ of the mixture \cite{Guo71}
(for the derivation of the theoretical mobility at a finite frequency, see Supplementary Material \cite{suppl}; in the theoretical mobility, the contribution of the electron scattering by thermally excited ripplons, which is not negligible at a high mobility, is also included).
As shown in Fig. \ref{mu_T_with_theory}, the theoretical mobility is in excellent agreement with the experimental data at high $x_3$ and high $T$ (except for $x_3 =$ 1.0\%, where the mean-free-path of bulk QPs is larger than the period of the DL) without any adjustable parameters.
This agreement suggests that there is no contribution from the 2D $^3$He layer in the viscous regime.

\begin{figure}
\begin{center}
\includegraphics[width=0.95\linewidth,keepaspectratio]{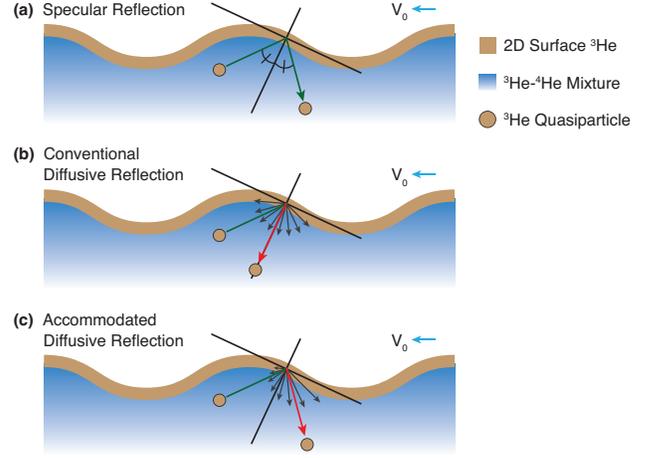}
\end{center}
\caption{\label{reflection} Schematic pictures of microscopic reflection processes of a $^3$He QP at the surface, seen in the reference frame moving horizontally together with the DL.
(a) Specular reflection. The $^3$He QP is reflected with an angle equal to the incident angle.
(b) Conventional diffusive reflection. Small arrows indicate the probability of QP reflection in a given direction. Red arrow indicates the averaged direction of reflection.
After reflection, $^3$He QPs are in thermal equilibrium with the moving surface.
(c) Accommodated diffusive reflection. Reflected QPs are in thermal equilibrium with the 2D $^3$He layer which is not moving horizontally with the DL. The drag force in this process is by a factor less than that of specular reflection (see text).
}
\end{figure}

The experimental mobility deviates from the viscosity-limited one and approaches a temperature-independent value at low temperatures.
These observations suggest a crossover from the viscous regime to the ballistic regime with decreasing $T$ as the mean free path $l_q$ of a $^3$He QP in the bulk mixture becomes longer than the lattice constant of the WC which is about  1 $\mu$m (in the Fermi liquid, $l_q$ increases with decreasing $T$ as $l_q \propto T^{-2}$). The mobility is then limited by friction caused by bulk QP reflection from the moving DL.
In the case of {\it pure} $^3$He, a $^3$He QP is demonstrated to be reflected specularly \cite{Shirahama97,Ikegami06,Monarkha1997}.
For the mixture, a similar process is shown in Fig.~\ref{reflection}(a); however, it is not trivial how a $^3$He QP is reflected from a surface element $dS$ in the presence of the 2D $^3$He layer.

As a reflection law, specular and diffusive reflections have been conventionally considered.
For the specular reflection of a $^3$He QP, the incident and reflection angles are equal [Fig. \ref{reflection}(a)].
In this case, there is no momentum transfer in the tangential direction to the surface element $dS$.
For the conventional diffusive reflection [Fig. \ref{reflection}(b)], which generally occurs at a solid surface with microscopic irregularities, reflected $^3$He QPs are in thermal equilibrium with the moving surface, and therefore their momentum distribution is significantly different from that of bulk QPs, resulting in a large average momentum exchange in the direction of motion.
Drag forces $dF_D$ acting on a surface element $dS$ thus differ by orders of magnitude for the two cases:
\begin{equation}
 dF_D  \sim n_3 p_F V_0 (\delta /a)^2 dS 
\end{equation}
for the specular reflection and
\begin{equation}
dF_D \sim n_3 p_F V_0 dS 
\end{equation}
for the conventional diffusive reflection \cite{Monarkha2006}, where $V_0$ is the horizontal velocity of the surface profile, $n_3$ and $p_F$ are the number density and the Fermi momentum of $^3$He in the mixture, $\delta \sim$ 0.1 \AA \ is the depth of the DL, and $a$($=$ 0.93 $\mu$m) is the lattice constant of the WC.
These suggest that
(i) for specular reflection, $dF_D$ is by $(\delta /a)^2$ ($\sim$ 10$^{-10}$) smaller than the case of  conventional diffusive reflection
and (ii) when $\delta \to 0$, $dF_D \to 0$ for specular reflection, while $dF_D$ is independent of the dimple depth for conventional diffusive reflection, which means that it is not applicable for the description of the drag force of the DL~\cite{Monarkha2006}.

As shown in Fig. \ref{mu_T_with_theory}, the theoretical mobility of the specular reflection model evaluated at our measurement frequency (black dashed line) is in qualitative agreement with the experimental mobility (see Supplementary Material for the derivation of the mobility at a nonzero frequency \cite{suppl}). However, the experimental mobility is still higher than that given by this model (as noted above, the conventional diffusive reflection results in  even much lower WC mobility).
This noteworthy result means that conventional specular and diffusive reflection laws cannot explain observed mobility data of the long mean-free-path regime.

As an unusual QP reflection model, Monarkha and Kono proposed a process involving the accommodation of an incoming $^3$He QP with the surface layer of $^3$He atoms  \cite{Monarkha2006}, which is shown schematically in Fig. \ref{reflection}(c).
Note that Fig. \ref{reflection} is drawn for an observer moving horizontally with the DL.
The key features of this process are (1) the momentum distribution of reflected QPs is described by the Fermi function $f_{0}\left( \varepsilon _{\beta ,p}\right)$ in the reference frame bound to the element $dS^\prime$ of the 2D $^3$He layer  (here $\varepsilon _{\beta ,p}$ is the energy of a QP with a momentum $\mathbf{p}$ and spin $\beta$), i.e. reflected QPs are in thermal equilibrium with the $^3$He layer,
and (2) the 2D $^3$He layer does not move horizontally together with the DL but just oscillates vertically with the amplitude of $\delta$ (this is the reason for a prime symbol in  $dS^\prime$ ).
Because of (1), this process represents diffusive reflection; when the DL is stationary the momentum distribution of reflected QPs is the same as that of the conventional diffusive model.
However, because of (2), the momentum distribution of reflected QPs is described by the equilibrium function in the frame which is not moving horizontally with a surface relief. 
Therefore, in the reference frame fixed to the DL, as qualitatively drawn in Fig.~\ref{reflection}(c), the momentum distribution function of reflected QPs 
\begin{equation}
f_{\mathrm{out}}\left( \mathbf{p}\right) =f_{0}\left( \varepsilon _{\beta ,p}+\mathbf{pV}_{0}+p_{z}\nabla \xi \mathbf{V}_{0}\right)
\label{e1}
\end{equation}
is close to the distribution function of incoming QPs
$f_{\mathrm{in}}\left( \mathbf{p}\right) =f_{0}\left( \varepsilon _{\beta ,p}+\mathbf{pV}_{0}\right) $, where $\xi (\mathbf{r})$ describes the surface relief of the DL. 
Thus, this reflection process reduces the average in-plane momentum exchange at the surface significantly. 
Only a small drag force is caused by the last term in the argument of the distribution function in Eq.~(\ref{e1}) associated with the oscillating vertical motion of the $^3$He layer with a small velocity of $\nabla \xi \mathbf{V}_{0} \simeq (\delta/a) V_0$.
Obviously, for a flat surface ($\nabla \xi = 0$), the drag force $F_{D}=0$.

Remarkably, such a simple modification of the diffusive reflection model leads to a giant increase in the WC mobility $-$ the drag sforce acting on the moving DL becomes smaller by a factor $(\delta /a)^2$ as compared to that given by conventional diffusive reflection.
Detailed theoretical analysis predicts that dc mobility is by a factor of four larger than that found for the case of specular reflection~\cite{Monarkha2006} and by a smaller factor at a finite frequency as shown with blue dashed lines in Fig.~\ref{mu_T_with_theory} (see Supplementary Material for the finite-frequency effect \cite{suppl}).
However, not all incoming QPs are reflected by this accommodated diffusive process; thus, we fit the data using the partial accommodation model (green solid lines in Fig.~\ref{mu_T_with_theory}), where a fraction $r_a$ among QPs are scattered by the accommodated diffusive process and the others
are reflected specularly. 
In this case, the dimple drag force is defined as
\begin{equation}
F_{D}=\left( 1-\frac{3}{4}r_{a} \right) F_{D}^{\left( \mathrm{spec}\right) } ,
\label{e2}
\end{equation}
where $F_{D}^{\left( \mathrm{spec}\right) }$ is the drag force for the specular reflection.

\begin{figure}[t]
\begin{center}
\includegraphics[width=0.65\linewidth,keepaspectratio]{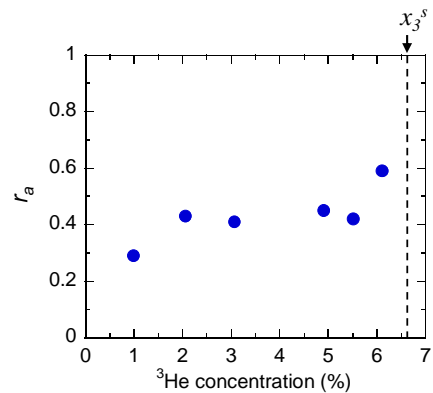}
\end{center}
\caption{\label{accommo_ratio} Accommodation ratio $r_a$ as a function of $^3$He concentration. 
}
\end{figure}

As can be seen in Fig.~\ref{accommo_ratio}, $r_a$ increases with increasing $x_3$.
This increase could be associated with a momentum mismatch between $^3$He in the surface layer and $^3$He in the bulk mixture caused by the large difference in the Fermi energies of the two systems: $T_F^{\rm 3D} <$ 0.4 K while $T_F^{\rm 2D} \sim$ 2 K [Fig. \ref{WC_and_surface_layer}(c)].
The mismatch suggests that the conservation of momentum and energy cannot be satisfied in the reflection process without involving other excitations such as surface waves, prohibiting the accommodation process.
The momentum mismatch becomes less significant at higher $x_3$, making the accommodation process more favorable.
Another possibility for the increase of $r_a$ with $x_3$ is associated with the increase of the thickness of the $^3$He layer according to 
Fig.~\ref{WC_and_surface_layer}(b).

Our work can be extended to lower temperatures where the 2D $^3$He system is expected to undergo a superfluid transition, potentially to a topological superfluid state.
The results obtained in this work indicate that the WC mobility could be useful for detecting the superfluid state of the surface layer because the transition should affect microscopic details of the accommodation process.

In conclusion, we have demonstrated that the interplay of the 2D and 3D Fermi systems generates a new kind of QP reflection from an uneven surface relief moving along the interface, which reveals itself as an anomalous enhancement in the WC mobility on the surface of $^3$He-$^4$He mixture.
%This study \textcolor{blue}{paves the way} to understanding general aspects of the dynamics of mixed-dimensional systems.
%\textcolor{red}{coexisting 2D and 3D Fermi liquids}.

%mixed-dimensional systems.

This work was partly supported by JSPS KAKENHI Grant Numbers JP24000007, JP26287084, and JP17H01145 and by National Research Foundation (NRF) of Korea Grant Numbers NRF-2015R1C1A1A01055813 and 2016R1A5A1008184.

%\cite{Kerr1964,Tanaka2000,LaudauLifshitz,Sherlock1973,Seligmann1969,Atkins1953,Eckardt1974,Saitoh77,Mehrotra1984,Monarkha1976,Monarkha1983,Monarkha2005}

\vspace{3mm}

%%%%%%%%%%%%%%%%%%%%%%%%%%%%%%%%%%%%%%%%%%%%%%%%%%%%%%%%%%%%%%%%%%%%%%%%%%%%

%\bibliography{mixture.bib}

\begin{thebibliography}{20}%
\makeatletter
\providecommand \@ifxundefined [1]{%
 \@ifx{#1\undefined}
}%
\providecommand \@ifnum [1]{%
 \ifnum #1\expandafter \@firstoftwo
 \else \expandafter \@secondoftwo
 \fi
}%
\providecommand \@ifx [1]{%
 \ifx #1\expandafter \@firstoftwo
 \else \expandafter \@secondoftwo
 \fi
}%
\providecommand \natexlab [1]{#1}%
\providecommand \enquote  [1]{``#1''}%
\providecommand \bibnamefont  [1]{#1}%
\providecommand \bibfnamefont [1]{#1}%
\providecommand \citenamefont [1]{#1}%
\providecommand \href@noop [0]{\@secondoftwo}%
\providecommand \href [0]{\begingroup \@sanitize@url \@href}%
\providecommand \@href[1]{\@@startlink{#1}\@@href}%
\providecommand \@@href[1]{\endgroup#1\@@endlink}%
\providecommand \@sanitize@url [0]{\catcode `\\12\catcode `\$12\catcode
  `\&12\catcode `\#12\catcode `\^12\catcode `\_12\catcode `\%12\relax}%
\providecommand \@@startlink[1]{}%
\providecommand \@@endlink[0]{}%
\providecommand \url  [0]{\begingroup\@sanitize@url \@url }%
\providecommand \@url [1]{\endgroup\@href {#1}{\urlprefix }}%
\providecommand \urlprefix  [0]{URL }%
\providecommand \Eprint [0]{\href }%
\providecommand \doibase [0]{http://dx.doi.org/}%
\providecommand \selectlanguage [0]{\@gobble}%
\providecommand \bibinfo  [0]{\@secondoftwo}%
\providecommand \bibfield  [0]{\@secondoftwo}%
\providecommand \translation [1]{[#1]}%
\providecommand \BibitemOpen [0]{}%
\providecommand \bibitemStop [0]{}%
\providecommand \bibitemNoStop [0]{.\EOS\space}%
\providecommand \EOS [0]{\spacefactor3000\relax}%
\providecommand \BibitemShut  [1]{\csname bibitem#1\endcsname}%
\let\auto@bib@innerbib\@empty
%</preamble>
\bibitem [{\citenamefont {Edwards}\ \emph {et~al.}(1969)\citenamefont
  {Edwards}, \citenamefont {Ifft},\ and\ \citenamefont
  {Sarwinski}}]{Edwards1969}%
  \BibitemOpen
  \bibfield  {author} {\bibinfo {author} {\bibfnamefont {D.~O.}\ \bibnamefont
  {Edwards}}, \bibinfo {author} {\bibfnamefont {E.~M.}\ \bibnamefont {Ifft}}, \
  and\ \bibinfo {author} {\bibfnamefont {R.~E.}\ \bibnamefont {Sarwinski}},\
  }\href@noop {} {\bibfield  {journal} {\bibinfo  {journal} {Phys. Rev.}\
  }\textbf {\bibinfo {volume} {177}},\ \bibinfo {pages} {380} (\bibinfo {year}
  {1969})}\BibitemShut {NoStop}%
\bibitem [{\citenamefont {Kerr}\ and\ \citenamefont {Taylor}(1964)}]{Kerr1964}%
  \BibitemOpen
  \bibfield  {author} {\bibinfo {author} {\bibfnamefont {E.~C.}\ \bibnamefont
  {Kerr}}\ and\ \bibinfo {author} {\bibfnamefont {R.}~\bibnamefont {Taylor}},\
  }\href@noop {} {\bibfield  {journal} {\bibinfo  {journal} {Ann. Phys.}\
  }\textbf {\bibinfo {volume} {26}},\ \bibinfo {pages} {292} (\bibinfo {year}
  {1964})}\BibitemShut {NoStop}%
\bibitem [{\citenamefont {Tanaka}\ \emph {et~al.}(2000)\citenamefont {Tanaka},
  \citenamefont {Hatakeyama}, \citenamefont {Noma},\ and\ \citenamefont
  {Satoh}}]{Tanaka2000}%
  \BibitemOpen
  \bibfield  {author} {\bibinfo {author} {\bibfnamefont {E.}~\bibnamefont
  {Tanaka}}, \bibinfo {author} {\bibfnamefont {K.}~\bibnamefont {Hatakeyama}},
  \bibinfo {author} {\bibfnamefont {S.}~\bibnamefont {Noma}}, \ and\ \bibinfo
  {author} {\bibfnamefont {T.}~\bibnamefont {Satoh}},\ }\href {\doibase
  http://dx.doi.org/10.1016/S0011-2275(00)00052-7} {\bibfield  {journal}
  {\bibinfo  {journal} {Cryogenics}\ }\textbf {\bibinfo {volume} {40}},\
  \bibinfo {pages} {365} (\bibinfo {year} {2000})}\BibitemShut {NoStop}%
\bibitem [{\citenamefont {Kuenhold}\ \emph {et~al.}(1972)\citenamefont
  {Kuenhold}, \citenamefont {Crum},\ and\ \citenamefont
  {Sarwinski}}]{Kuenhold1972}%
  \BibitemOpen
  \bibfield  {author} {\bibinfo {author} {\bibfnamefont {K.~A.}\ \bibnamefont
  {Kuenhold}}, \bibinfo {author} {\bibfnamefont {D.~B.}\ \bibnamefont {Crum}},
  \ and\ \bibinfo {author} {\bibfnamefont {R.~E.}\ \bibnamefont {Sarwinski}},\
  }\href@noop {} {\bibfield  {journal} {\bibinfo  {journal} {Phys. Lett. A}\
  }\textbf {\bibinfo {volume} {41}},\ \bibinfo {pages} {13} (\bibinfo {year}
  {1972})}\BibitemShut {NoStop}%
\bibitem [{\citenamefont {K\"onig}\ and\ \citenamefont
  {Pobell}(1994)}]{Konig1994}%
  \BibitemOpen
  \bibfield  {author} {\bibinfo {author} {\bibfnamefont {R.}~\bibnamefont
  {K\"onig}}\ and\ \bibinfo {author} {\bibfnamefont {F.}~\bibnamefont
  {Pobell}},\ }\href@noop {} {\bibfield  {journal} {\bibinfo  {journal} {J. Low
  Temp. Phys.}\ }\textbf {\bibinfo {volume} {97}},\ \bibinfo {pages} {287}
  (\bibinfo {year} {1994})}\BibitemShut {NoStop}%
\bibitem [{\citenamefont {Guo}\ \emph {et~al.}(1971)\citenamefont {Guo},
  \citenamefont {Edwards}, \citenamefont {Sarwinski},\ and\ \citenamefont
  {Tough}}]{Guo1971}%
  \BibitemOpen
  \bibfield  {author} {\bibinfo {author} {\bibfnamefont {H.~M.}\ \bibnamefont
  {Guo}}, \bibinfo {author} {\bibfnamefont {D.~O.}\ \bibnamefont {Edwards}},
  \bibinfo {author} {\bibfnamefont {R.~E.}\ \bibnamefont {Sarwinski}}, \ and\
  \bibinfo {author} {\bibfnamefont {J.~T.}\ \bibnamefont {Tough}},\ }\href
  {\doibase 10.1103/PhysRevLett.27.1259} {\bibfield  {journal} {\bibinfo
  {journal} {Phys. Rev. Lett.}\ }\textbf {\bibinfo {volume} {27}},\ \bibinfo
  {pages} {1259} (\bibinfo {year} {1971})}\BibitemShut {NoStop}%
\bibitem [{\citenamefont {Landau}\ and\ \citenamefont
  {Lifshitz}(1980)}]{LaudauLifshitz}%
  \BibitemOpen
  \bibfield  {author} {\bibinfo {author} {\bibfnamefont {L.~D.}\ \bibnamefont
  {Landau}}\ and\ \bibinfo {author} {\bibfnamefont {E.~M.}\ \bibnamefont
  {Lifshitz}},\ }\href@noop {} {\emph {\bibinfo {title} {Statistical Physics,
  Part 1, 3rd Edition}}}\ (\bibinfo  {publisher} {Butterworth-Heinemann},\
  \bibinfo {year} {1980})\ Chap.~\bibinfo {chapter} {XV}\BibitemShut {NoStop}%
\bibitem [{\citenamefont {Sherlock}\ and\ \citenamefont
  {Edwards}(1973)}]{Sherlock}%
  \BibitemOpen
  \bibfield  {author} {\bibinfo {author} {\bibfnamefont {R.~A.}\ \bibnamefont
  {Sherlock}}\ and\ \bibinfo {author} {\bibfnamefont {D.~O.}\ \bibnamefont
  {Edwards}},\ }\href {\doibase 10.1103/PhysRevA.8.2744} {\bibfield  {journal}
  {\bibinfo  {journal} {Phys. Rev. A}\ }\textbf {\bibinfo {volume} {8}},\
  \bibinfo {pages} {2744} (\bibinfo {year} {1973})}\BibitemShut {NoStop}%
\bibitem [{\citenamefont {Seligmann}\ \emph {et~al.}(1969)\citenamefont
  {Seligmann}, \citenamefont {Edwards}, \citenamefont {Sarwinski},\ and\
  \citenamefont {Tough}}]{Seligmann1969}%
  \BibitemOpen
  \bibfield  {author} {\bibinfo {author} {\bibfnamefont {P.}~\bibnamefont
  {Seligmann}}, \bibinfo {author} {\bibfnamefont {D.~O.}\ \bibnamefont
  {Edwards}}, \bibinfo {author} {\bibfnamefont {R.~E.}\ \bibnamefont
  {Sarwinski}}, \ and\ \bibinfo {author} {\bibfnamefont {J.~T.}\ \bibnamefont
  {Tough}},\ }\href {\doibase 10.1103/PhysRev.181.415} {\bibfield  {journal}
  {\bibinfo  {journal} {Phys. Rev.}\ }\textbf {\bibinfo {volume} {181}},\
  \bibinfo {pages} {415} (\bibinfo {year} {1969})}\BibitemShut {NoStop}%
\bibitem [{\citenamefont {Atkins}(1953)}]{Atkins1953}%
  \BibitemOpen
  \bibfield  {author} {\bibinfo {author} {\bibfnamefont {K.~R.}\ \bibnamefont
  {Atkins}},\ }\href {\doibase 10.1139/p53-101} {\bibfield  {journal} {\bibinfo
   {journal} {Can. J. Phys.}\ }\textbf {\bibinfo {volume} {31}},\ \bibinfo
  {pages} {1165} (\bibinfo {year} {1953})}\BibitemShut {NoStop}%
\bibitem [{\citenamefont {Eckardt}\ \emph {et~al.}(1974)\citenamefont
  {Eckardt}, \citenamefont {Edwards}, \citenamefont {Fatouros}, \citenamefont
  {Gasparini},\ and\ \citenamefont {Shen}}]{Eckardt1974}%
  \BibitemOpen
  \bibfield  {author} {\bibinfo {author} {\bibfnamefont {J.~R.}\ \bibnamefont
  {Eckardt}}, \bibinfo {author} {\bibfnamefont {D.~O.}\ \bibnamefont
  {Edwards}}, \bibinfo {author} {\bibfnamefont {P.~P.}\ \bibnamefont
  {Fatouros}}, \bibinfo {author} {\bibfnamefont {F.~M.}\ \bibnamefont
  {Gasparini}}, \ and\ \bibinfo {author} {\bibfnamefont {S.~Y.}\ \bibnamefont
  {Shen}},\ }\href {\doibase 10.1103/PhysRevLett.32.706} {\bibfield  {journal}
  {\bibinfo  {journal} {Phys. Rev. Lett.}\ }\textbf {\bibinfo {volume} {32}},\
  \bibinfo {pages} {706} (\bibinfo {year} {1974})}\BibitemShut {NoStop}%
\bibitem [{\citenamefont {Monarkha}\ and\ \citenamefont
  {Kono}(2004)}]{BookMonarkha}%
  \BibitemOpen
  \bibfield  {author} {\bibinfo {author} {\bibfnamefont {Y.~P.}\ \bibnamefont
  {Monarkha}}\ and\ \bibinfo {author} {\bibfnamefont {K.}~\bibnamefont
  {Kono}},\ }\href@noop {} {\emph {\bibinfo {title} {Two-Dimensional Coulomb
  Liquids and Solids}}}\ (\bibinfo  {publisher} {Springer-Verlag},\ \bibinfo
  {address} {Berlin},\ \bibinfo {year} {2004})\BibitemShut {NoStop}%
\bibitem [{\citenamefont {Saitoh}(1977)}]{Saitoh77}%
  \BibitemOpen
  \bibfield  {author} {\bibinfo {author} {\bibfnamefont {M.}~\bibnamefont
  {Saitoh}},\ }\href@noop {} {\bibfield  {journal} {\bibinfo  {journal} {J.
  Phys. Soc. Jpn.}\ }\textbf {\bibinfo {volume} {42}},\ \bibinfo {pages} {201}
  (\bibinfo {year} {1977})}\BibitemShut {NoStop}%
\bibitem [{\citenamefont {Mehrotra}\ \emph {et~al.}(1984)\citenamefont
  {Mehrotra}, \citenamefont {Guo}, \citenamefont {Ruan}, \citenamefont {Mast},\
  and\ \citenamefont {Dahm}}]{Mehrotra1984}%
  \BibitemOpen
  \bibfield  {author} {\bibinfo {author} {\bibfnamefont {R.}~\bibnamefont
  {Mehrotra}}, \bibinfo {author} {\bibfnamefont {C.~J.}\ \bibnamefont {Guo}},
  \bibinfo {author} {\bibfnamefont {Y.~Z.}\ \bibnamefont {Ruan}}, \bibinfo
  {author} {\bibfnamefont {D.~B.}\ \bibnamefont {Mast}}, \ and\ \bibinfo
  {author} {\bibfnamefont {A.~J.}\ \bibnamefont {Dahm}},\ }\href@noop {}
  {\bibfield  {journal} {\bibinfo  {journal} {Phys. Rev. B}\ }\textbf {\bibinfo
  {volume} {29}},\ \bibinfo {pages} {5239} (\bibinfo {year}
  {1984})}\BibitemShut {NoStop}%
\bibitem [{\citenamefont {Monarkha}\ and\ \citenamefont
  {Kono}(1997)}]{Monarkha1997}%
  \BibitemOpen
  \bibfield  {author} {\bibinfo {author} {\bibfnamefont {Y.~P.}\ \bibnamefont
  {Monarkha}}\ and\ \bibinfo {author} {\bibfnamefont {K.}~\bibnamefont
  {Kono}},\ }\href@noop {} {\bibfield  {journal} {\bibinfo  {journal} {J. Phys.
  Soc. Jpn.}\ }\textbf {\bibinfo {volume} {66}},\ \bibinfo {pages} {3901}
  (\bibinfo {year} {1997})}\BibitemShut {NoStop}%
\bibitem [{\citenamefont {Monarkha}\ and\ \citenamefont
  {Syvokon}(2012)}]{Monarkha2012}%
  \BibitemOpen
  \bibfield  {author} {\bibinfo {author} {\bibfnamefont {Y.~P.}\ \bibnamefont
  {Monarkha}}\ and\ \bibinfo {author} {\bibfnamefont {V.~E.}\ \bibnamefont
  {Syvokon}},\ }\href@noop {} {\bibfield  {journal} {\bibinfo  {journal} {Low
  Temp. Phys.}\ }\textbf {\bibinfo {volume} {38}},\ \bibinfo {pages} {1067}
  (\bibinfo {year} {2012})}\BibitemShut {NoStop}%
\bibitem [{\citenamefont {Monarkha}(1976)}]{Monarkha1976}%
  \BibitemOpen
  \bibfield  {author} {\bibinfo {author} {\bibfnamefont {Y.~P.}\ \bibnamefont
  {Monarkha}},\ }\href@noop {} {\bibfield  {journal} {\bibinfo  {journal} {Sov.
  J. Low Temp. Phys.}\ }\textbf {\bibinfo {volume} {2}},\ \bibinfo {pages}
  {600} (\bibinfo {year} {1976})}\BibitemShut {NoStop}%
\bibitem [{\citenamefont {Monarkha}\ and\ \citenamefont
  {Shikin}(1983)}]{Monarkha1983}%
  \BibitemOpen
  \bibfield  {author} {\bibinfo {author} {\bibfnamefont {Y.~P.}\ \bibnamefont
  {Monarkha}}\ and\ \bibinfo {author} {\bibfnamefont {V.~B.}\ \bibnamefont
  {Shikin}},\ }\href@noop {} {\bibfield  {journal} {\bibinfo  {journal} {Sov.
  J. Low Temp. Phys.}\ }\textbf {\bibinfo {volume} {9}},\ \bibinfo {pages}
  {471} (\bibinfo {year} {1983})}\BibitemShut {NoStop}%
\bibitem [{\citenamefont {Monarkha}\ and\ \citenamefont
  {Kono}(2005)}]{Monarkha2005}%
  \BibitemOpen
  \bibfield  {author} {\bibinfo {author} {\bibfnamefont {Y.~P.}\ \bibnamefont
  {Monarkha}}\ and\ \bibinfo {author} {\bibfnamefont {K.}~\bibnamefont
  {Kono}},\ }\href@noop {} {\bibfield  {journal} {\bibinfo  {journal} {J. Phys.
  Soc. Jpn.}\ }\textbf {\bibinfo {volume} {74}},\ \bibinfo {pages} {960}
  (\bibinfo {year} {2005})}\BibitemShut {NoStop}%
\bibitem [{\citenamefont {Monarkha}\ and\ \citenamefont
  {Kono}(2006)}]{Monarkha2006}%
  \BibitemOpen
  \bibfield  {author} {\bibinfo {author} {\bibfnamefont {Y.~P.}\ \bibnamefont
  {Monarkha}}\ and\ \bibinfo {author} {\bibfnamefont {K.}~\bibnamefont
  {Kono}},\ }\href@noop {} {\bibfield  {journal} {\bibinfo  {journal} {J. Phys.
  Soc. Jpn.}\ }\textbf {\bibinfo {volume} {75}},\ \bibinfo {pages} {044601}
  (\bibinfo {year} {2006})}\BibitemShut {NoStop}%
\end{thebibliography}%


\begin{thebibliography}{40}%
\makeatletter
\providecommand \@ifxundefined [1]{%
 \@ifx{#1\undefined}
}%
\providecommand \@ifnum [1]{%
 \ifnum #1\expandafter \@firstoftwo
 \else \expandafter \@secondoftwo
 \fi
}%
\providecommand \@ifx [1]{%
 \ifx #1\expandafter \@firstoftwo
 \else \expandafter \@secondoftwo
 \fi
}%
\providecommand \natexlab [1]{#1}%
\providecommand \enquote  [1]{``#1''}%
\providecommand \bibnamefont  [1]{#1}%
\providecommand \bibfnamefont [1]{#1}%
\providecommand \citenamefont [1]{#1}%
\providecommand \href@noop [0]{\@secondoftwo}%
\providecommand \href [0]{\begingroup \@sanitize@url \@href}%
\providecommand \@href[1]{\@@startlink{#1}\@@href}%
\providecommand \@@href[1]{\endgroup#1\@@endlink}%
\providecommand \@sanitize@url [0]{\catcode `\\12\catcode `\$12\catcode
  `\&12\catcode `\#12\catcode `\^12\catcode `\_12\catcode `\%12\relax}%
\providecommand \@@startlink[1]{}%
\providecommand \@@endlink[0]{}%
\providecommand \url  [0]{\begingroup\@sanitize@url \@url }%
\providecommand \@url [1]{\endgroup\@href {#1}{\urlprefix }}%
\providecommand \urlprefix  [0]{URL }%
\providecommand \Eprint [0]{\href }%
\providecommand \doibase [0]{http://dx.doi.org/}%
\providecommand \selectlanguage [0]{\@gobble}%
\providecommand \bibinfo  [0]{\@secondoftwo}%
\providecommand \bibfield  [0]{\@secondoftwo}%
\providecommand \translation [1]{[#1]}%
\providecommand \BibitemOpen [0]{}%
\providecommand \bibitemStop [0]{}%
\providecommand \bibitemNoStop [0]{.\EOS\space}%
\providecommand \EOS [0]{\spacefactor3000\relax}%
\providecommand \BibitemShut  [1]{\csname bibitem#1\endcsname}%
\let\auto@bib@innerbib\@empty
%</preamble>
\bibitem [{\citenamefont {Zangwill}(1988)}]{book_Zangwill}%
  \BibitemOpen
  \bibfield  {author} {\bibinfo {author} {\bibfnamefont {A.}~\bibnamefont
  {Zangwill}},\ }\href@noop {} {\emph {\bibinfo {title} {Physics at
  Surfaces}}}\ (\bibinfo  {publisher} {Cambridge University Press},\ \bibinfo
  {address} {Cambridge},\ \bibinfo {year} {1988})\BibitemShut {NoStop}%
\bibitem [{\citenamefont {Plumb}\ \emph {et~al.}(2014)\citenamefont {Plumb},
  \citenamefont {Salluzzo}, \citenamefont {Razzoli}, \citenamefont
  {M\aa{}nsson}, \citenamefont {Falub}, \citenamefont {Krempasky},
  \citenamefont {Matt}, \citenamefont {Chang}, \citenamefont {Schulte},
  \citenamefont {Braun}, \citenamefont {Ebert}, \citenamefont {Min\'ar},
  \citenamefont {Delley}, \citenamefont {Zhou}, \citenamefont {Schmitt},
  \citenamefont {Shi}, \citenamefont {Mesot}, \citenamefont {Patthey},\ and\
  \citenamefont {Radovi\ifmmode~\acute{c}\else \'{c}\fi{}}}]{Plumb2014}%
  \BibitemOpen
  \bibfield  {author} {\bibinfo {author} {\bibfnamefont {N.~C.}\ \bibnamefont
  {Plumb}}, \bibinfo {author} {\bibfnamefont {M.}~\bibnamefont {Salluzzo}},
  \bibinfo {author} {\bibfnamefont {E.}~\bibnamefont {Razzoli}}, \bibinfo
  {author} {\bibfnamefont {M.}~\bibnamefont {M\aa{}nsson}}, \bibinfo {author}
  {\bibfnamefont {M.}~\bibnamefont {Falub}}, \bibinfo {author} {\bibfnamefont
  {J.}~\bibnamefont {Krempasky}}, \bibinfo {author} {\bibfnamefont {C.~E.}\
  \bibnamefont {Matt}}, \bibinfo {author} {\bibfnamefont {J.}~\bibnamefont
  {Chang}}, \bibinfo {author} {\bibfnamefont {M.}~\bibnamefont {Schulte}},
  \bibinfo {author} {\bibfnamefont {J.}~\bibnamefont {Braun}}, \bibinfo
  {author} {\bibfnamefont {H.}~\bibnamefont {Ebert}}, \bibinfo {author}
  {\bibfnamefont {J.}~\bibnamefont {Min\'ar}}, \bibinfo {author} {\bibfnamefont
  {B.}~\bibnamefont {Delley}}, \bibinfo {author} {\bibfnamefont {K.-J.}\
  \bibnamefont {Zhou}}, \bibinfo {author} {\bibfnamefont {T.}~\bibnamefont
  {Schmitt}}, \bibinfo {author} {\bibfnamefont {M.}~\bibnamefont {Shi}},
  \bibinfo {author} {\bibfnamefont {J.}~\bibnamefont {Mesot}}, \bibinfo
  {author} {\bibfnamefont {L.}~\bibnamefont {Patthey}}, \ and\ \bibinfo
  {author} {\bibfnamefont {M.}~\bibnamefont {Radovi\ifmmode~\acute{c}\else
  \'{c}\fi{}}},\ }\href {\doibase 10.1103/PhysRevLett.113.086801} {\bibfield
  {journal} {\bibinfo  {journal} {Phys. Rev. Lett.}\ }\textbf {\bibinfo
  {volume} {113}},\ \bibinfo {pages} {086801} (\bibinfo {year}
  {2014})}\BibitemShut {NoStop}%
\bibitem [{\citenamefont {Wang}\ \emph {et~al.}(2012)\citenamefont {Wang},
  \citenamefont {Hsieh}, \citenamefont {Sie}, \citenamefont {Steinberg},
  \citenamefont {Gardner}, \citenamefont {Lee}, \citenamefont
  {Jarillo-Herrero},\ and\ \citenamefont {Gedik}}]{Wang2012}%
  \BibitemOpen
  \bibfield  {author} {\bibinfo {author} {\bibfnamefont {Y.~H.}\ \bibnamefont
  {Wang}}, \bibinfo {author} {\bibfnamefont {D.}~\bibnamefont {Hsieh}},
  \bibinfo {author} {\bibfnamefont {E.~J.}\ \bibnamefont {Sie}}, \bibinfo
  {author} {\bibfnamefont {H.}~\bibnamefont {Steinberg}}, \bibinfo {author}
  {\bibfnamefont {D.~R.}\ \bibnamefont {Gardner}}, \bibinfo {author}
  {\bibfnamefont {Y.~S.}\ \bibnamefont {Lee}}, \bibinfo {author} {\bibfnamefont
  {P.}~\bibnamefont {Jarillo-Herrero}}, \ and\ \bibinfo {author} {\bibfnamefont
  {N.}~\bibnamefont {Gedik}},\ }\href {\doibase 10.1103/PhysRevLett.109.127401}
  {\bibfield  {journal} {\bibinfo  {journal} {Phys. Rev. Lett.}\ }\textbf
  {\bibinfo {volume} {109}},\ \bibinfo {pages} {127401} (\bibinfo {year}
  {2012})}\BibitemShut {NoStop}%
\bibitem [{\citenamefont {Nishida}\ and\ \citenamefont
  {Tan}(2008)}]{Nishida2008}%
  \BibitemOpen
  \bibfield  {author} {\bibinfo {author} {\bibfnamefont {Y.}~\bibnamefont
  {Nishida}}\ and\ \bibinfo {author} {\bibfnamefont {S.}~\bibnamefont {Tan}},\
  }\href {\doibase 10.1103/PhysRevLett.101.170401} {\bibfield  {journal}
  {\bibinfo  {journal} {Phys. Rev. Lett.}\ }\textbf {\bibinfo {volume} {101}},\
  \bibinfo {pages} {170401} (\bibinfo {year} {2008})}\BibitemShut {NoStop}%
\bibitem [{\citenamefont {Iskin}\ and\ \citenamefont
  {Suba\c{s}\i}(2010)}]{Iskin2010}%
  \BibitemOpen
  \bibfield  {author} {\bibinfo {author} {\bibfnamefont {M.}~\bibnamefont
  {Iskin}}\ and\ \bibinfo {author} {\bibfnamefont {A.~L.}\ \bibnamefont
  {Suba\c{s}\i}},\ }\href {\doibase 10.1103/PhysRevA.82.063628} {\bibfield
  {journal} {\bibinfo  {journal} {Phys. Rev. A}\ }\textbf {\bibinfo {volume}
  {82}},\ \bibinfo {pages} {063628} (\bibinfo {year} {2010})}\BibitemShut
  {NoStop}%
\bibitem [{\citenamefont {Okamoto}\ \emph {et~al.}(2017)\citenamefont
  {Okamoto}, \citenamefont {Mathey},\ and\ \citenamefont
  {Huang}}]{Okamoto2017}%
  \BibitemOpen
  \bibfield  {author} {\bibinfo {author} {\bibfnamefont {J.}~\bibnamefont
  {Okamoto}}, \bibinfo {author} {\bibfnamefont {L.}~\bibnamefont {Mathey}}, \
  and\ \bibinfo {author} {\bibfnamefont {W.-M.}\ \bibnamefont {Huang}},\
  }\href@noop {} {\bibfield  {journal} {\bibinfo  {journal} {Phys. Rev. A}\
  }\textbf {\bibinfo {volume} {95}},\ \bibinfo {pages} {053633} (\bibinfo
  {year} {2017})}\BibitemShut {NoStop}%
\bibitem [{\citenamefont {Dobbs}(2000)}]{Dobbs00}%
  \BibitemOpen
  \bibfield  {author} {\bibinfo {author} {\bibfnamefont {E.~R.}\ \bibnamefont
  {Dobbs}},\ }\href@noop {} {\emph {\bibinfo {title} {Helium Three}}}\
  (\bibinfo  {publisher} {Oxford University Press},\ \bibinfo {address}
  {Oxford},\ \bibinfo {year} {2000})\BibitemShut {NoStop}%
\bibitem [{\citenamefont {Guo}\ \emph {et~al.}(1971)\citenamefont {Guo},
  \citenamefont {Edwards}, \citenamefont {Sarwinski},\ and\ \citenamefont
  {Tough}}]{Guo71}%
  \BibitemOpen
  \bibfield  {author} {\bibinfo {author} {\bibfnamefont {H.~M.}\ \bibnamefont
  {Guo}}, \bibinfo {author} {\bibfnamefont {D.~O.}\ \bibnamefont {Edwards}},
  \bibinfo {author} {\bibfnamefont {R.~E.}\ \bibnamefont {Sarwinski}}, \ and\
  \bibinfo {author} {\bibfnamefont {J.~T.}\ \bibnamefont {Tough}},\ }\href@noop
  {} {\bibfield  {journal} {\bibinfo  {journal} {Phys. Rev. Lett.}\ }\textbf
  {\bibinfo {volume} {27}},\ \bibinfo {pages} {1259} (\bibinfo {year}
  {1971})}\BibitemShut {NoStop}%
\bibitem [{\citenamefont {Edwards}\ \emph {et~al.}(1975)\citenamefont
  {Edwards}, \citenamefont {Shen}, \citenamefont {Eckardt}, \citenamefont
  {Fatouros},\ and\ \citenamefont {Gasparini}}]{Edwards75}%
  \BibitemOpen
  \bibfield  {author} {\bibinfo {author} {\bibfnamefont {D.~O.}\ \bibnamefont
  {Edwards}}, \bibinfo {author} {\bibfnamefont {S.~Y.}\ \bibnamefont {Shen}},
  \bibinfo {author} {\bibfnamefont {J.~R.}\ \bibnamefont {Eckardt}}, \bibinfo
  {author} {\bibfnamefont {P.~P.}\ \bibnamefont {Fatouros}}, \ and\ \bibinfo
  {author} {\bibfnamefont {F.~M.}\ \bibnamefont {Gasparini}},\ }\href@noop {}
  {\bibfield  {journal} {\bibinfo  {journal} {Phys. Rev. B}\ }\textbf {\bibinfo
  {volume} {12}},\ \bibinfo {pages} {892} (\bibinfo {year} {1975})}\BibitemShut
  {NoStop}%
\bibitem [{\citenamefont {Yorozu}\ \emph {et~al.}(1992)\citenamefont {Yorozu},
  \citenamefont {Hiroi}, \citenamefont {Fukuyama}, \citenamefont {Akimoto},
  \citenamefont {Ishimoto},\ and\ \citenamefont {Ogawa}}]{Yorozu1992}%
  \BibitemOpen
  \bibfield  {author} {\bibinfo {author} {\bibfnamefont {S.}~\bibnamefont
  {Yorozu}}, \bibinfo {author} {\bibfnamefont {M.}~\bibnamefont {Hiroi}},
  \bibinfo {author} {\bibfnamefont {H.}~\bibnamefont {Fukuyama}}, \bibinfo
  {author} {\bibfnamefont {H.}~\bibnamefont {Akimoto}}, \bibinfo {author}
  {\bibfnamefont {H.}~\bibnamefont {Ishimoto}}, \ and\ \bibinfo {author}
  {\bibfnamefont {S.}~\bibnamefont {Ogawa}},\ }\href@noop {} {\bibfield
  {journal} {\bibinfo  {journal} {Phys. Rev. B}\ }\textbf {\bibinfo {volume}
  {45}},\ \bibinfo {pages} {12942} (\bibinfo {year} {1992})}\BibitemShut
  {NoStop}%
\bibitem [{\citenamefont {Andreev}(1966)}]{Andreev1966}%
  \BibitemOpen
  \bibfield  {author} {\bibinfo {author} {\bibfnamefont {A.~F.}\ \bibnamefont
  {Andreev}},\ }\href@noop {} {\bibfield  {journal} {\bibinfo  {journal} {Sov.
  Phys. JETP}\ }\textbf {\bibinfo {volume} {23}},\ \bibinfo {pages} {939}
  (\bibinfo {year} {1966})}\BibitemShut {NoStop}%
\bibitem [{\citenamefont {Saam}(1971)}]{Saam1971}%
  \BibitemOpen
  \bibfield  {author} {\bibinfo {author} {\bibfnamefont {W.~F.}\ \bibnamefont
  {Saam}},\ }\href {\doibase 10.1103/PhysRevA.4.1278} {\bibfield  {journal}
  {\bibinfo  {journal} {Phys. Rev. A}\ }\textbf {\bibinfo {volume} {4}},\
  \bibinfo {pages} {1278} (\bibinfo {year} {1971})}\BibitemShut {NoStop}%
\bibitem [{Rev()}]{Review_ProgLowTempPhys}%
  \BibitemOpen
  \href@noop {} {}\bibinfo {note} {D. O. Edwards and W. F. Saam, The Free
  Surface of Liquid Helium. {\it Progress in Low Temperature Physics, Vol. 7,
  Part A}, Chap. 4, edited by Brewer, D (North-Holland, 1978) .}\BibitemShut
  {Stop}%
\bibitem [{\citenamefont {Andrei}(1997)}]{Book2DE}%
  \BibitemOpen
  \bibinfo {editor} {\bibfnamefont {E.~Y.}\ \bibnamefont {Andrei}},\ ed.,\
  \href@noop {} {\emph {\bibinfo {title} {Two-Dimensional Electron Systems on
  Helium and Other Cryogenic Substrates}}}\ (\bibinfo  {publisher} {Kluwer
  Academic Publishers},\ \bibinfo {address} {Dordrecht},\ \bibinfo {year}
  {1997})\BibitemShut {NoStop}%
\bibitem [{\citenamefont {Monarkha}\ and\ \citenamefont
  {Kono}(2004)}]{BookMonarkha}%
  \BibitemOpen
  \bibfield  {author} {\bibinfo {author} {\bibfnamefont {Y.~P.}\ \bibnamefont
  {Monarkha}}\ and\ \bibinfo {author} {\bibfnamefont {K.}~\bibnamefont
  {Kono}},\ }\href@noop {} {\emph {\bibinfo {title} {Two-Dimensional Coulomb
  Liquids and Solids}}}\ (\bibinfo  {publisher} {Springer-Verlag},\ \bibinfo
  {address} {Berlin},\ \bibinfo {year} {2004})\BibitemShut {NoStop}%
\bibitem [{\citenamefont {Deville}(1988)}]{Deville1988}%
  \BibitemOpen
  \bibfield  {author} {\bibinfo {author} {\bibfnamefont {G.}~\bibnamefont
  {Deville}},\ }\href {\doibase 10.1007/BF00681729} {\bibfield  {journal}
  {\bibinfo  {journal} {Journal of Low Temperature Physics}\ }\textbf {\bibinfo
  {volume} {72}},\ \bibinfo {pages} {135} (\bibinfo {year} {1988})}\BibitemShut
  {NoStop}%
\bibitem [{\citenamefont {Shirahama}\ \emph {et~al.}(1997)\citenamefont
  {Shirahama}, \citenamefont {Kirichek},\ and\ \citenamefont
  {Kono}}]{Shirahama97}%
  \BibitemOpen
  \bibfield  {author} {\bibinfo {author} {\bibfnamefont {K.}~\bibnamefont
  {Shirahama}}, \bibinfo {author} {\bibfnamefont {O.~I.}\ \bibnamefont
  {Kirichek}}, \ and\ \bibinfo {author} {\bibfnamefont {K.}~\bibnamefont
  {Kono}},\ }\href@noop {} {\bibfield  {journal} {\bibinfo  {journal} {Phys.
  Rev. Lett.}\ }\textbf {\bibinfo {volume} {79}},\ \bibinfo {pages} {4218}
  (\bibinfo {year} {1997})}\BibitemShut {NoStop}%
\bibitem [{\citenamefont {Ikegami}\ and\ \citenamefont
  {Kono}(2006)}]{Ikegami06}%
  \BibitemOpen
  \bibfield  {author} {\bibinfo {author} {\bibfnamefont {H.}~\bibnamefont
  {Ikegami}}\ and\ \bibinfo {author} {\bibfnamefont {K.}~\bibnamefont {Kono}},\
  }\href@noop {} {\bibfield  {journal} {\bibinfo  {journal} {Phys. Rev. Lett.}\
  }\textbf {\bibinfo {volume} {97}},\ \bibinfo {pages} {165303} (\bibinfo
  {year} {2006})}\BibitemShut {NoStop}%
\bibitem [{\citenamefont {Monarkha}\ and\ \citenamefont
  {Kono}(1997)}]{Monarkha1997}%
  \BibitemOpen
  \bibfield  {author} {\bibinfo {author} {\bibfnamefont {Y.~P.}\ \bibnamefont
  {Monarkha}}\ and\ \bibinfo {author} {\bibfnamefont {K.}~\bibnamefont
  {Kono}},\ }\href@noop {} {\bibfield  {journal} {\bibinfo  {journal} {J. Phys.
  Soc. Jpn.}\ }\textbf {\bibinfo {volume} {66}},\ \bibinfo {pages} {3901}
  (\bibinfo {year} {1997})}\BibitemShut {NoStop}%
\bibitem [{\citenamefont {Yayama}\ and\ \citenamefont
  {Yatsuyama}(2014)}]{Yayama_JLTP2014}%
  \BibitemOpen
  \bibfield  {author} {\bibinfo {author} {\bibfnamefont {H.}~\bibnamefont
  {Yayama}}\ and\ \bibinfo {author} {\bibfnamefont {Y.}~\bibnamefont
  {Yatsuyama}},\ }\href {\doibase 10.1007/s10909-013-0956-9} {\bibfield
  {journal} {\bibinfo  {journal} {J. Low Temp. Phys.}\ }\textbf {\bibinfo
  {volume} {175}},\ \bibinfo {pages} {401} (\bibinfo {year}
  {2014})}\BibitemShut {NoStop}%
\bibitem [{sup()}]{suppl}%
  \BibitemOpen
  \href@noop {} {}\bibinfo {note} {See Supplemental Material at [url] for
  details on the thickness and the Fermi energy of the surface $^3$He layer,
  the mobility of electrons in ripplon scattering regime, and the
  finite-frequency effect on the mobility, which includes Refs.
  [29-40].}\BibitemShut {Stop}%
\bibitem [{\citenamefont {Sommer}\ and\ \citenamefont
  {Tanner}(1971)}]{Sommer71}%
  \BibitemOpen
  \bibfield  {author} {\bibinfo {author} {\bibfnamefont {W.~T.}\ \bibnamefont
  {Sommer}}\ and\ \bibinfo {author} {\bibfnamefont {D.~J.}\ \bibnamefont
  {Tanner}},\ }\href@noop {} {\bibfield  {journal} {\bibinfo  {journal} {Phys.
  Rev. Lett.}\ }\textbf {\bibinfo {volume} {27}},\ \bibinfo {pages} {1345}
  (\bibinfo {year} {1971})}\BibitemShut {NoStop}%
\bibitem [{\citenamefont {Monarkha}\ and\ \citenamefont
  {Syvokon}(2012)}]{Monarkha2012}%
  \BibitemOpen
  \bibfield  {author} {\bibinfo {author} {\bibfnamefont {Y.~P.}\ \bibnamefont
  {Monarkha}}\ and\ \bibinfo {author} {\bibfnamefont {V.~E.}\ \bibnamefont
  {Syvokon}},\ }\href@noop {} {\bibfield  {journal} {\bibinfo  {journal} {Low
  Temp. Phys.}\ }\textbf {\bibinfo {volume} {38}},\ \bibinfo {pages} {1067}
  (\bibinfo {year} {2012})}\BibitemShut {NoStop}%
\bibitem [{\citenamefont {Kuenhold}\ \emph {et~al.}(1972)\citenamefont
  {Kuenhold}, \citenamefont {Crum},\ and\ \citenamefont
  {Sarwinski}}]{Kuenhold1972}%
  \BibitemOpen
  \bibfield  {author} {\bibinfo {author} {\bibfnamefont {K.~A.}\ \bibnamefont
  {Kuenhold}}, \bibinfo {author} {\bibfnamefont {D.~B.}\ \bibnamefont {Crum}},
  \ and\ \bibinfo {author} {\bibfnamefont {R.~E.}\ \bibnamefont {Sarwinski}},\
  }\href@noop {} {\bibfield  {journal} {\bibinfo  {journal} {Phys. Lett. A}\
  }\textbf {\bibinfo {volume} {41}},\ \bibinfo {pages} {13} (\bibinfo {year}
  {1972})}\BibitemShut {NoStop}%
\bibitem [{\citenamefont {{Van Der Boog}}\ \emph {et~al.}(1981)\citenamefont
  {{Van Der Boog}}, \citenamefont {Husson}, \citenamefont {Disatnik},\ and\
  \citenamefont {Kramers}}]{VanDerBoog1981}%
  \BibitemOpen
  \bibfield  {author} {\bibinfo {author} {\bibfnamefont {A.~G.~M.}\
  \bibnamefont {{Van Der Boog}}}, \bibinfo {author} {\bibfnamefont {L.~P.~J.}\
  \bibnamefont {Husson}}, \bibinfo {author} {\bibfnamefont {Y.}~\bibnamefont
  {Disatnik}}, \ and\ \bibinfo {author} {\bibfnamefont {H.~C.}\ \bibnamefont
  {Kramers}},\ }\href@noop {} {\bibfield  {journal} {\bibinfo  {journal}
  {Physica B+C}\ }\textbf {\bibinfo {volume} {104}},\ \bibinfo {pages} {303}
  (\bibinfo {year} {1981})}\BibitemShut {NoStop}%
\bibitem [{\citenamefont {K\"onig}\ and\ \citenamefont
  {Pobell}(1994)}]{Konig1994}%
  \BibitemOpen
  \bibfield  {author} {\bibinfo {author} {\bibfnamefont {R.}~\bibnamefont
  {K\"onig}}\ and\ \bibinfo {author} {\bibfnamefont {F.}~\bibnamefont
  {Pobell}},\ }\href@noop {} {\bibfield  {journal} {\bibinfo  {journal} {J. Low
  Temp. Phys.}\ }\textbf {\bibinfo {volume} {97}},\ \bibinfo {pages} {287}
  (\bibinfo {year} {1994})}\BibitemShut {NoStop}%
\bibitem [{\citenamefont {Edwards}\ \emph {et~al.}(1969)\citenamefont
  {Edwards}, \citenamefont {Ifft},\ and\ \citenamefont
  {Sarwinski}}]{Edwards1969}%
  \BibitemOpen
  \bibfield  {author} {\bibinfo {author} {\bibfnamefont {D.~O.}\ \bibnamefont
  {Edwards}}, \bibinfo {author} {\bibfnamefont {E.~M.}\ \bibnamefont {Ifft}}, \
  and\ \bibinfo {author} {\bibfnamefont {R.~E.}\ \bibnamefont {Sarwinski}},\
  }\href@noop {} {\bibfield  {journal} {\bibinfo  {journal} {Phys. Rev.}\
  }\textbf {\bibinfo {volume} {177}},\ \bibinfo {pages} {380} (\bibinfo {year}
  {1969})}\BibitemShut {NoStop}%
\bibitem [{\citenamefont {Monarkha}\ and\ \citenamefont
  {Kono}(2006)}]{Monarkha2006}%
  \BibitemOpen
  \bibfield  {author} {\bibinfo {author} {\bibfnamefont {Y.~P.}\ \bibnamefont
  {Monarkha}}\ and\ \bibinfo {author} {\bibfnamefont {K.}~\bibnamefont
  {Kono}},\ }\href@noop {} {\bibfield  {journal} {\bibinfo  {journal} {J. Phys.
  Soc. Jpn.}\ }\textbf {\bibinfo {volume} {75}},\ \bibinfo {pages} {044601}
  (\bibinfo {year} {2006})}\BibitemShut {NoStop}%
\bibitem [{\citenamefont {Kerr}\ and\ \citenamefont {Taylor}(1964)}]{Kerr1964}%
  \BibitemOpen
  \bibfield  {author} {\bibinfo {author} {\bibfnamefont {E.~C.}\ \bibnamefont
  {Kerr}}\ and\ \bibinfo {author} {\bibfnamefont {R.}~\bibnamefont {Taylor}},\
  }\href@noop {} {\bibfield  {journal} {\bibinfo  {journal} {Ann. Phys.}\
  }\textbf {\bibinfo {volume} {26}},\ \bibinfo {pages} {292} (\bibinfo {year}
  {1964})}\BibitemShut {NoStop}%
\bibitem [{\citenamefont {Tanaka}\ \emph {et~al.}(2000)\citenamefont {Tanaka},
  \citenamefont {Hatakeyama}, \citenamefont {Noma},\ and\ \citenamefont
  {Satoh}}]{Tanaka2000}%
  \BibitemOpen
  \bibfield  {author} {\bibinfo {author} {\bibfnamefont {E.}~\bibnamefont
  {Tanaka}}, \bibinfo {author} {\bibfnamefont {K.}~\bibnamefont {Hatakeyama}},
  \bibinfo {author} {\bibfnamefont {S.}~\bibnamefont {Noma}}, \ and\ \bibinfo
  {author} {\bibfnamefont {T.}~\bibnamefont {Satoh}},\ }\href {\doibase
  http://dx.doi.org/10.1016/S0011-2275(00)00052-7} {\bibfield  {journal}
  {\bibinfo  {journal} {Cryogenics}\ }\textbf {\bibinfo {volume} {40}},\
  \bibinfo {pages} {365} (\bibinfo {year} {2000})}\BibitemShut {NoStop}%
\bibitem [{\citenamefont {Landau}\ and\ \citenamefont
  {Lifshitz}(1980)}]{LaudauLifshitz}%
  \BibitemOpen
  \bibfield  {author} {\bibinfo {author} {\bibfnamefont {L.~D.}\ \bibnamefont
  {Landau}}\ and\ \bibinfo {author} {\bibfnamefont {E.~M.}\ \bibnamefont
  {Lifshitz}},\ }\href@noop {} {\emph {\bibinfo {title} {Statistical Physics,
  Part 1, 3rd Edition}}}\ (\bibinfo  {publisher} {Butterworth-Heinemann},\
  \bibinfo {year} {1980})\ Chap.~\bibinfo {chapter} {XV}\BibitemShut {NoStop}%
\bibitem [{\citenamefont {Sherlock}\ and\ \citenamefont
  {Edwards}(1973)}]{Sherlock1973}%
  \BibitemOpen
  \bibfield  {author} {\bibinfo {author} {\bibfnamefont {R.~A.}\ \bibnamefont
  {Sherlock}}\ and\ \bibinfo {author} {\bibfnamefont {D.~O.}\ \bibnamefont
  {Edwards}},\ }\href {\doibase 10.1103/PhysRevA.8.2744} {\bibfield  {journal}
  {\bibinfo  {journal} {Phys. Rev. A}\ }\textbf {\bibinfo {volume} {8}},\
  \bibinfo {pages} {2744} (\bibinfo {year} {1973})}\BibitemShut {NoStop}%
\bibitem [{\citenamefont {Seligmann}\ \emph {et~al.}(1969)\citenamefont
  {Seligmann}, \citenamefont {Edwards}, \citenamefont {Sarwinski},\ and\
  \citenamefont {Tough}}]{Seligmann1969}%
  \BibitemOpen
  \bibfield  {author} {\bibinfo {author} {\bibfnamefont {P.}~\bibnamefont
  {Seligmann}}, \bibinfo {author} {\bibfnamefont {D.~O.}\ \bibnamefont
  {Edwards}}, \bibinfo {author} {\bibfnamefont {R.~E.}\ \bibnamefont
  {Sarwinski}}, \ and\ \bibinfo {author} {\bibfnamefont {J.~T.}\ \bibnamefont
  {Tough}},\ }\href {\doibase 10.1103/PhysRev.181.415} {\bibfield  {journal}
  {\bibinfo  {journal} {Phys. Rev.}\ }\textbf {\bibinfo {volume} {181}},\
  \bibinfo {pages} {415} (\bibinfo {year} {1969})}\BibitemShut {NoStop}%
\bibitem [{\citenamefont {Atkins}(1953)}]{Atkins1953}%
  \BibitemOpen
  \bibfield  {author} {\bibinfo {author} {\bibfnamefont {K.~R.}\ \bibnamefont
  {Atkins}},\ }\href {\doibase 10.1139/p53-101} {\bibfield  {journal} {\bibinfo
   {journal} {Can. J. Phys.}\ }\textbf {\bibinfo {volume} {31}},\ \bibinfo
  {pages} {1165} (\bibinfo {year} {1953})}\BibitemShut {NoStop}%
\bibitem [{\citenamefont {Eckardt}\ \emph {et~al.}(1974)\citenamefont
  {Eckardt}, \citenamefont {Edwards}, \citenamefont {Fatouros}, \citenamefont
  {Gasparini},\ and\ \citenamefont {Shen}}]{Eckardt1974}%
  \BibitemOpen
  \bibfield  {author} {\bibinfo {author} {\bibfnamefont {J.~R.}\ \bibnamefont
  {Eckardt}}, \bibinfo {author} {\bibfnamefont {D.~O.}\ \bibnamefont
  {Edwards}}, \bibinfo {author} {\bibfnamefont {P.~P.}\ \bibnamefont
  {Fatouros}}, \bibinfo {author} {\bibfnamefont {F.~M.}\ \bibnamefont
  {Gasparini}}, \ and\ \bibinfo {author} {\bibfnamefont {S.~Y.}\ \bibnamefont
  {Shen}},\ }\href {\doibase 10.1103/PhysRevLett.32.706} {\bibfield  {journal}
  {\bibinfo  {journal} {Phys. Rev. Lett.}\ }\textbf {\bibinfo {volume} {32}},\
  \bibinfo {pages} {706} (\bibinfo {year} {1974})}\BibitemShut {NoStop}%
\bibitem [{\citenamefont {Saitoh}(1977)}]{Saitoh77}%
  \BibitemOpen
  \bibfield  {author} {\bibinfo {author} {\bibfnamefont {M.}~\bibnamefont
  {Saitoh}},\ }\href@noop {} {\bibfield  {journal} {\bibinfo  {journal} {J.
  Phys. Soc. Jpn.}\ }\textbf {\bibinfo {volume} {42}},\ \bibinfo {pages} {201}
  (\bibinfo {year} {1977})}\BibitemShut {NoStop}%
\bibitem [{\citenamefont {Mehrotra}\ \emph {et~al.}(1984)\citenamefont
  {Mehrotra}, \citenamefont {Guo}, \citenamefont {Ruan}, \citenamefont {Mast},\
  and\ \citenamefont {Dahm}}]{Mehrotra1984}%
  \BibitemOpen
  \bibfield  {author} {\bibinfo {author} {\bibfnamefont {R.}~\bibnamefont
  {Mehrotra}}, \bibinfo {author} {\bibfnamefont {C.~J.}\ \bibnamefont {Guo}},
  \bibinfo {author} {\bibfnamefont {Y.~Z.}\ \bibnamefont {Ruan}}, \bibinfo
  {author} {\bibfnamefont {D.~B.}\ \bibnamefont {Mast}}, \ and\ \bibinfo
  {author} {\bibfnamefont {A.~J.}\ \bibnamefont {Dahm}},\ }\href@noop {}
  {\bibfield  {journal} {\bibinfo  {journal} {Phys. Rev. B}\ }\textbf {\bibinfo
  {volume} {29}},\ \bibinfo {pages} {5239} (\bibinfo {year}
  {1984})}\BibitemShut {NoStop}%
\bibitem [{\citenamefont {Monarkha}(1976)}]{Monarkha1976}%
  \BibitemOpen
  \bibfield  {author} {\bibinfo {author} {\bibfnamefont {Y.~P.}\ \bibnamefont
  {Monarkha}},\ }\href@noop {} {\bibfield  {journal} {\bibinfo  {journal} {Sov.
  J. Low Temp. Phys.}\ }\textbf {\bibinfo {volume} {2}},\ \bibinfo {pages}
  {600} (\bibinfo {year} {1976})}\BibitemShut {NoStop}%
\bibitem [{\citenamefont {Monarkha}\ and\ \citenamefont
  {Shikin}(1983)}]{Monarkha1983}%
  \BibitemOpen
  \bibfield  {author} {\bibinfo {author} {\bibfnamefont {Y.~P.}\ \bibnamefont
  {Monarkha}}\ and\ \bibinfo {author} {\bibfnamefont {V.~B.}\ \bibnamefont
  {Shikin}},\ }\href@noop {} {\bibfield  {journal} {\bibinfo  {journal} {Sov.
  J. Low Temp. Phys.}\ }\textbf {\bibinfo {volume} {9}},\ \bibinfo {pages}
  {471} (\bibinfo {year} {1983})}\BibitemShut {NoStop}%
\bibitem [{\citenamefont {Monarkha}\ and\ \citenamefont
  {Kono}(2005)}]{Monarkha2005}%
  \BibitemOpen
  \bibfield  {author} {\bibinfo {author} {\bibfnamefont {Y.~P.}\ \bibnamefont
  {Monarkha}}\ and\ \bibinfo {author} {\bibfnamefont {K.}~\bibnamefont
  {Kono}},\ }\href@noop {} {\bibfield  {journal} {\bibinfo  {journal} {J. Phys.
  Soc. Jpn.}\ }\textbf {\bibinfo {volume} {74}},\ \bibinfo {pages} {960}
  (\bibinfo {year} {2005})}\BibitemShut {NoStop}%
\end{thebibliography}


\end{document}

%\preprint{kkk}

\title{Supplementary Material: \\
Anomalous Quasiparticle Reflection \\ 
from the Surface of a $^3$He-$^4$He Dilute Solution}

% repeat the \author .. \affiliation  etc. as neededS

\author{Hiroki Ikegami}\email{hikegami@riken.jp}
\affiliation{RIKEN Center for Emergent Matter Science (CEMS), Wako, Saitama 351-0198, Japan}

\author{Kitak Kim}
\affiliation{Department of Physics, KAIST,  Daejeon 34141, Republic of Korea}

\author{Daisuke Sato}
\affiliation{RIKEN Center for Emergent Matter Science (CEMS), Wako, Saitama 351-0198, Japan}

\author{Kimitoshi Kono}
\affiliation{RIKEN Center for Emergent Matter Science (CEMS), Wako, Saitama 351-0198, Japan}

\author{Hyoungsoon Choi}\email{h.choi@kaist.ac.kr}
\affiliation{Department of Physics, KAIST,  Daejeon 34141, Republic of Korea}

\author{Yuriy P. Monarkha}
\affiliation{Institute for Low Temperature Physics and Engineering, 47 Nauky Avenue, Kharkov 61103, Ukraine}

%\thanks{u}

\date{\today}

\maketitle

%********************************************************************

\section{Interpolation of Physical Quantities of $^3$He-$^4$He Mixture}
\label{interpolation}

In the main text, we use interpolated values of physical quantities of a $^3$He-$^4$He mixture to analyze the experimental data.
In this section, we describe how we interpolate these quantities.

\subsection{Molar Volume}

As the molar volume of the dilute mixture $V_m$, we adopt the results of Edwards {\it et al.} \cite{Edwards1969}:
\begin{equation}
V_m = V_4 \left[ {1 + \alpha \left( {x_3 ,T} \right)x_3 } \right]
\end{equation} 
with
\begin{equation}
\alpha = (0.284\pm 0.005)- \left[ (0.032\pm 0.003) \right] T.
\end{equation} 
Here $V_4=$ 27.579 cc/mol is the molar volume of pure $^4$He \cite{Kerr1964,Tanaka2000}, $x_3$ is the concentration of $^3$He, and $T$ is the temperature in the unit of K. 
This equation is valid for $^3$He concentrations less than 6.6
$\%$ and temperatures below 0.6 K.
Our estimates of the dielectric constant and the density of the mixture are also based on the molar volume obtained from this equation.

For the calculations of the theoretical mobility described in the main text, we neglect the temperature dependence of $V_m$ because $V_m$ changes by less than 0.06\% from 0 to 0.3 K at a fixed $x_3$.

\subsection{Viscosity}

The viscosity of the dilute mixture has been measured by Kuenhold {\it et al.} (for $x_3=$ 0.5, 1.3, 2.7, 5.0, and 7.0\%) \cite{Kuenhold1972} and K\"onig and Pobell ($x_3=$ 0.98 and 6.1\%) \cite{Konig1994}.
We interpolate these experimental data in the temperature range of 10$-$300 mK as a function of $T$ and $x_3$ according to the following procedure.
First, we divide the temperature range into two regions: above and below $T_0$, which varies between 50 and 65 mK depending on $x_3$.
Within each temperature region, we fit the viscosity $\eta$ to the following function of $T$ and $x_3$:
\begin{equation}
\ln (\eta T^2 ) = \sum\limits_{m = 0}^4 {\sum\limits_{n = 0}^1 {A_{mn} (\ln T)^m (\ln x_3 )^n } }, 
\label{viscosity_fit}
\end{equation}
yielding two sets of fitting parameters, $A_{mn}^{H}$ and $A_{mn}^L$, for the high- and low-temperature regions, respectively.

The two fitted curves in the two temperature regions are smoothly connected by imposing the relation:
\begin{equation}
\ln (\eta T^2 ) =f(T-T_0) \ln (\eta^H T^2 )  + [1- f(T-T_0)] \ln (\eta^L T^2 ) ,
\label{viscosity_connection}
\end{equation}
where $\eta^H$ and $\eta^L$ are the viscosity above and below $T_0$, respectively. We use the function $f(T)$ given by
\begin{equation}
 f(T)= \frac{1}{{1 + \exp ( - T/\alpha )}} 
\end{equation}
with $\alpha =$ 0.003 K.
Figure \ref{viscosity_fitting}(a) shows the results of our fitting. 
The interpolated values of viscosity for the values of $x_3$ used in our measurements are shown in Fig.~\ref{viscosity_fitting}(b). 
The error in our estimation of the viscosity is less than 7\%.

\renewcommand{\thefigure}{S1}
\begin{figure}
\begin{center}
\includegraphics[width=0.9\linewidth,keepaspectratio]{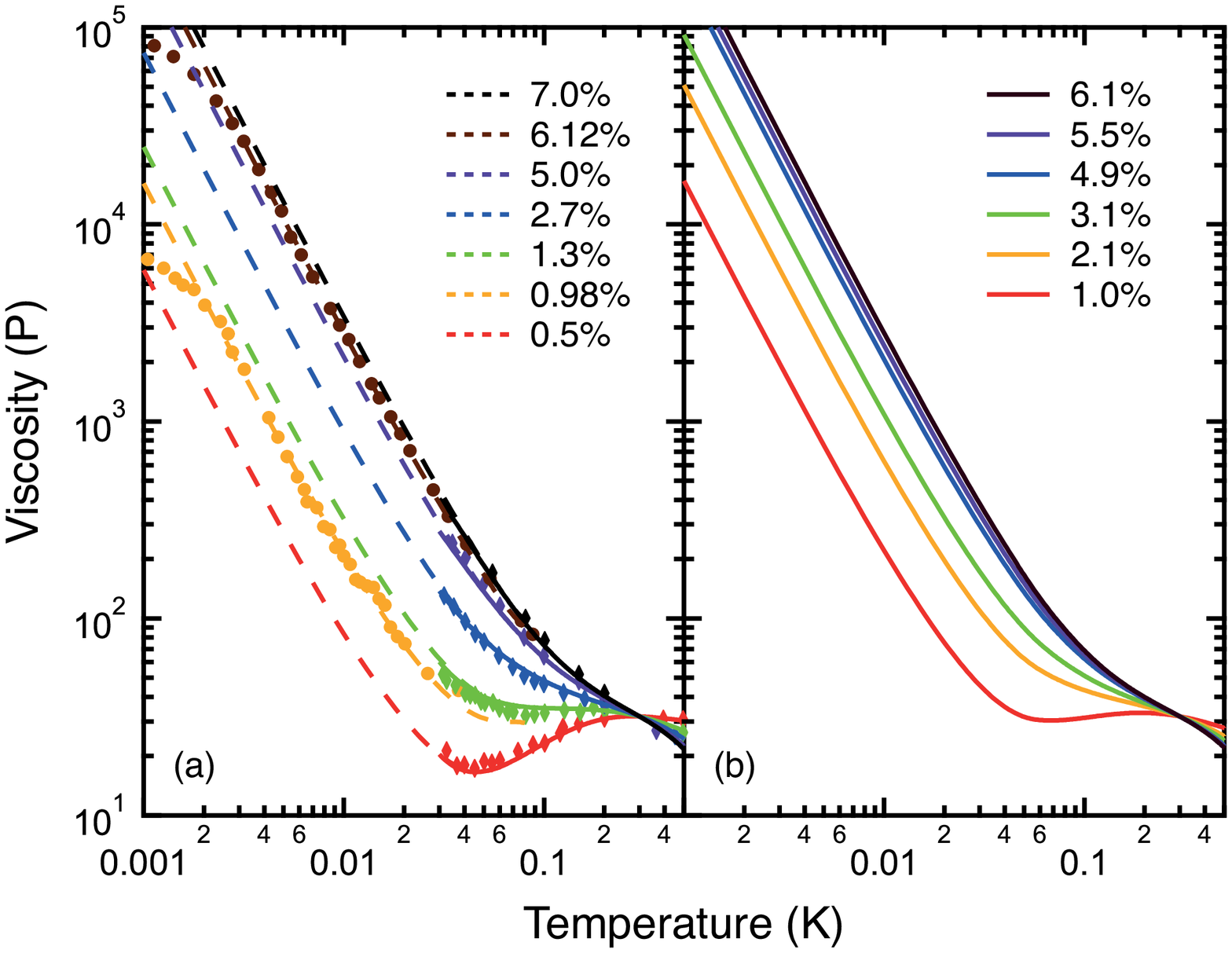}
\end{center}
\caption{\label{viscosity_fitting} (a) Fitting of the viscosity data of Kuenhold {\it et al.} \cite{Kuenhold1972} and K\"onig and Pobell \cite{Konig1994} by the procedure described in the supplemental text. (b) Interpolated viscosity at the $^3$He concentrations used in our experiment.
}
\vspace {10mm}
\end{figure}

\renewcommand{\thefigure}{S2}
\begin{figure}[t]
\begin{center}
\includegraphics[width=0.9\linewidth,keepaspectratio]{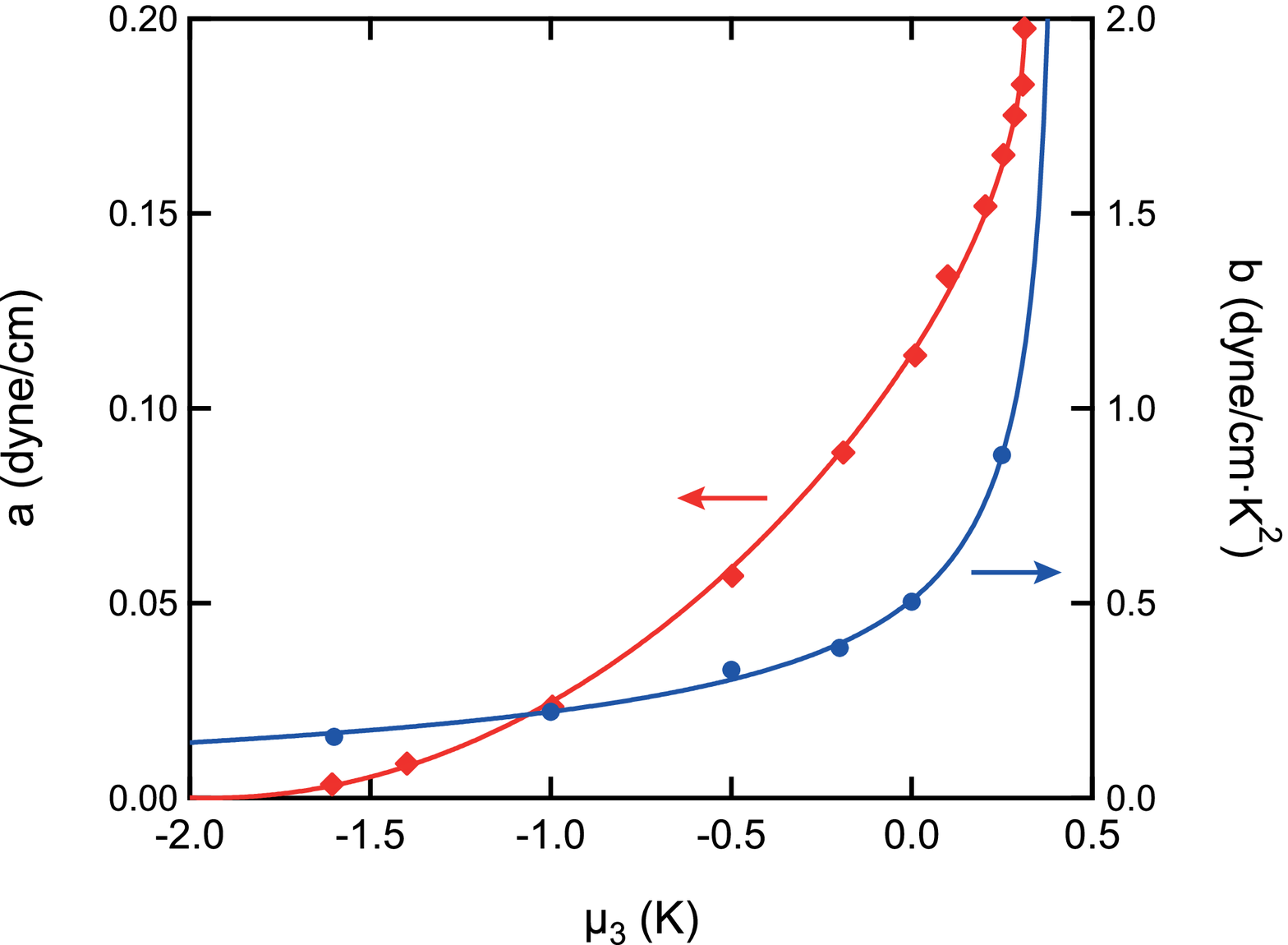}
\end{center}
\caption{\label{fitting_surface_tension} Results of fitting of $a$ and $b$ with Eqs. (\ref{fitparameter1}) and (\ref{fitparameter2}). The values of $a$ and $b$ are taken from Ref. \cite{Guo1971}.
}
\vspace {10mm}
\end{figure}

\subsection{Surface Tension}
\label{interpolation_sigma}

\renewcommand{\thefigure}{S3}
\begin{figure}[t]
\begin{center}
\includegraphics[width=0.9\linewidth,keepaspectratio]{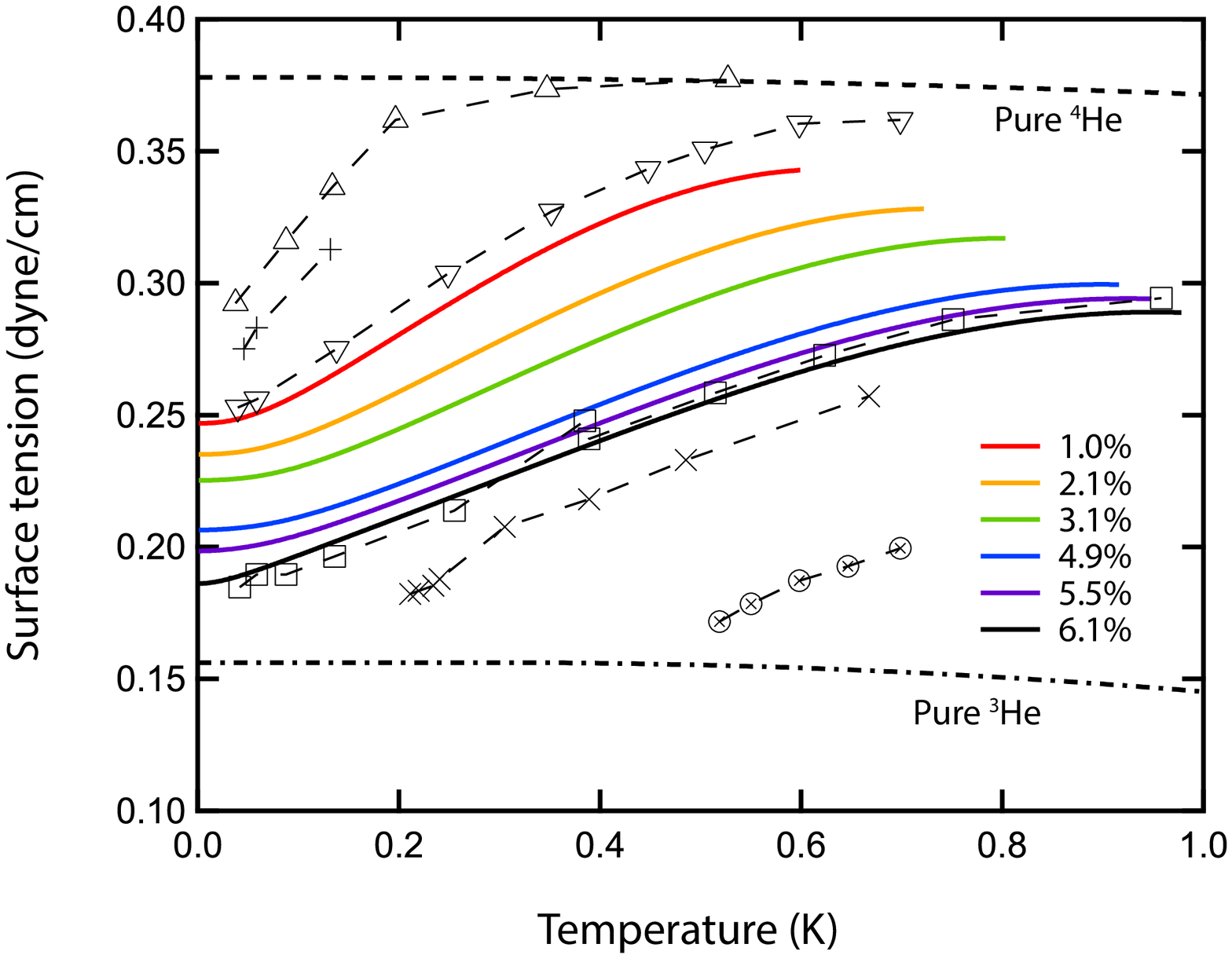}
\end{center}
\caption{\label{interpolation_surface_tension} Interpolated surface tension. The open symbols represent the experimental data of the surface tension for each $^3$He concentration, taken from Ref. \cite{Guo1971}: $\bigtriangleup$ 32 ppm, $+$ 542 ppm, $\bigtriangledown$ 0.56\%, $\square$ 6.16\%, $\times$ 9.6\%, $\otimes$ 22\%.
}
\vspace {10mm}
\end{figure}

The available experimental data on the surface tension of the dilute mixture are very limited, particularly in the range of $x_3=$ 0.5$-$6.1\%.
The data measured with the finest steps are those reported by Guo {\it et al.} ($x_3=$ 3.2$\times$10$^{-3}$, 5.42$\times$10$^{-2}$, 0.56, 6.16, 9.64, and 22\%) \cite{Guo1971}.
We interpolate their data using the following phenomenological model.
We assume that the surface layer of $^3$He is in thermal equilibrium with $^3$He atoms in the bulk having a chemical potential $\mu_3$.
According to the thermodynamic relation,  the surface entropy $S$ per unit area is related to the surface tension $\alpha _m$ as \cite{LaudauLifshitz}
\begin{equation}
S  = -\left( {\partial \alpha _m /\partial T} \right)_{\mu _3 }.
\label{entropy1}
\end{equation}
Here the surface entropy is composed of two parts, $S=S_3+S_4$, where $S_3$ is the entropy of $^3$He on the surface and $S_4$ is that of pure $^4$He.
$S_4$ and the surface tension of pure $^4$He $\alpha_4$ are related by $S_4 = - d\alpha_4 /dT$.
If $^3$He in the surface layer is Fermi degenerate, the entropy should be linear in temperature, $S_3=bT$, leading to
\begin{equation}
S = -\left( d\alpha_4 /dT \right) +bT.
\label{entropy2}
\end{equation}
From Eqs. (\ref{entropy1}) and (\ref{entropy2}), we derive
\begin{equation}
\Delta \alpha(T) \equiv \alpha_4 (T) - \alpha_m (T)  = a  + bT^2/2,
\label{sigma1}
\end{equation}
where $a$ and $b$ are functions of $\mu_3$.
We fit the experimental values of $a$ and $b$ obtained by Guo {\it et al.} \cite{Guo1971} as a function of $\mu_3$ to the following empirical forms: 
\begin{equation}
\sqrt{a} =-B\sqrt{\frac{(\mu _3-\alpha )^2}{A^2}-1} +\beta 
\label{fitparameter1}
\end{equation}
with fitting parameters $A$, $B$, $\alpha$, and $\beta$, and
\begin{equation}
b =C+\frac{D}{\sqrt{F-\mu _3}}
\label{fitparameter2}
\end{equation}
with fitting parameters $C$, $D$, and $F$.
The results of the fitting are shown in Fig. \ref{fitting_surface_tension}.

Next, we evaluate the dependence of $\mu_3$ on $x_3$ and $T$.  
The dependence is obtained by approximating $\mu_3$ as 
\begin{equation}
\mu_3 (T,x_3) \simeq  \mu_3 (0,x_3) + \mu_F (T,x_3), 
\label{chem_approx}
\end{equation}
where $\mu_F (T,x_3)$ is the chemical potential for a 3D non-interacting Fermi gas with effective mass $m_{\rm 3D}=$ 2.28$m_3$ \cite{Sherlock} ($m_3$ is the mass of a $^3$He atom).
For $\mu_3 (0,x_3)$, we use the experimental values obtained by Seligmann {\it et al.} \cite{Seligmann1969}.
From Eqs. (\ref{sigma1})$-$(\ref{chem_approx}) and using the temperature dependence of the experimental surface tension of pure $^4$He \cite{Guo1971,Atkins1953},
\begin{equation}
\alpha_4(T) = 0.387 - 0.0065 T^{7/3} \quad  \rm{(dyne/cm)}, 
\label{sigma_4}
\end{equation}
($T$ is in the unit of K), we obtain the interpolated $\alpha_m (T)$ as a function of $x_3$ and $T$.
The results of the interpolated $\alpha_m (T)$ at several fixed $x_3$ are shown in Fig. \ref{interpolation_surface_tension}.
The error in $\alpha_m (T)$ is estimated to be less than 2\%.

%\clearpage

%********************************************************************

\section{Thickness and Fermi Energy of Surface $^3$He Layer}
\label{surface_layer}

Here we consider how the areal density of the 2D $^3$He layer $n_s$ evolves with increasing $^3$He concentration in the bulk $x_3$.
According to  the thermodynamic relation, $n_s$ is described in terms of the surface tension $\alpha_m$ by
\begin{equation}
n_s=-\left. {\frac{{\partial \alpha_m}}{{\partial \mu _3 }}} \right|_T ,
\label{surface_layer_density}
\end{equation}
where $\mu_3$ is the chemical potential of $^3$He.
(Here we assume that the interaction between $^3$He in the surface layer and $^3$He in the bulk mixture does not depend on $\mu_3$.)
Using the dependence of $\alpha_m$ on $\mu _3$ obtained by Guo {\it et al.} [Eq. (\ref{sigma1})] \cite{Guo1971}, we calculate $n_s (\mu_3)$ from Eq. (\ref{surface_layer_density}). 
The chemical potential $\mu _3$ is approximately related to $x_3$ and $T$ by Eq. (\ref{chem_approx}).
Therefore, we obtain $n_s$ as a function of $x_3$ and $T$.
We show $n_s$ at $T=0$ as a function of $x_3$ in Fig. 1b of the main text. 
It increases with increasing $x_3$ and diverges as it approaches the saturation concentration of $^3$He.

Next, we consider how the Fermi energy of the 2D $^3$He layer $T_F^{\rm 2D}$ evolves with increasing $x_3$.
We consider that a potential well present near the surface generates bound states of $^3$He with eigenenergies of $E_{\bot}^n$ in the direction normal to the surface ($n=$ 0, 1, 2, $\cdots$, $l$).
The bound $^3$He can move freely with a kinetic energy $E_k^n$ in the plane parallel to the surface.
In this situation, the energy of the lowest bound state $E_{\bot}^0$ corresponds to the binding energy $E_b$.
Because the surface $^3$He layer is in equilibrium with bulk $^3$He in the mixture, the bound states with energies less than the Fermi energy of bulk $^3$He $E_F^{\rm 3D}$ are occupied at $T=$ 0. 
This leads to the relation $E_F^{\rm 2D}+E_b = E_F^{\rm 3D}$, where the Fermi energy of the 2D $^3$He layer $E_F^{\rm 2D}$ corresponds to the highest kinetic energy of the bound states of $n=$ 0.
For calculating $E_F^{\rm 2D}$ and $E_F^{\rm 3D}$ shown in Fig. 1c in the main text, we use the experimental value of Ref. \cite{Seligmann1969} for $E_F^{\rm 3D}$ and the experimental value of $E_b=-$ 2.28 K \cite{Eckardt1974}.

%\clearpage

%********************************************************************

\section{Mobility on pure $^4$He in Ripplon Scattering Regime}

The magnitude of the mobility presented in the main text is calibrated by multiplying by a factor about unity (0.94$-$1.14) so that the mobility at temperatures above $T_m$ agrees with the theoretical mobility of the highly correlated electrons in the ripplon scattering regime.
The justification for this calibration process is based on the fact that our mobility measured for pure $^4$He at temperatures above $T_m$ agrees with the theoretical mobility in both the ripplon [Eq. (3.70) in Ref. \cite{BookMonarkha}] and gas scattering \cite{Saitoh77} regimes  without correction of the magnitude of the mobility as shown in Fig. \ref{mobility_pure4He}.
Note that the mobility for pure $^4$He measured by Mehrotra {\it et al.} at similar electron densities deviates from the theoretical mobility in the ripplon scattering regime \cite{Mehrotra1984}.
The origin of the discrepancy between our and their data is unknown.

\vspace{10mm}

\renewcommand{\thefigure}{S4}
\begin{figure}[bht]
\begin{center}
\includegraphics[width=0.85\linewidth,keepaspectratio]{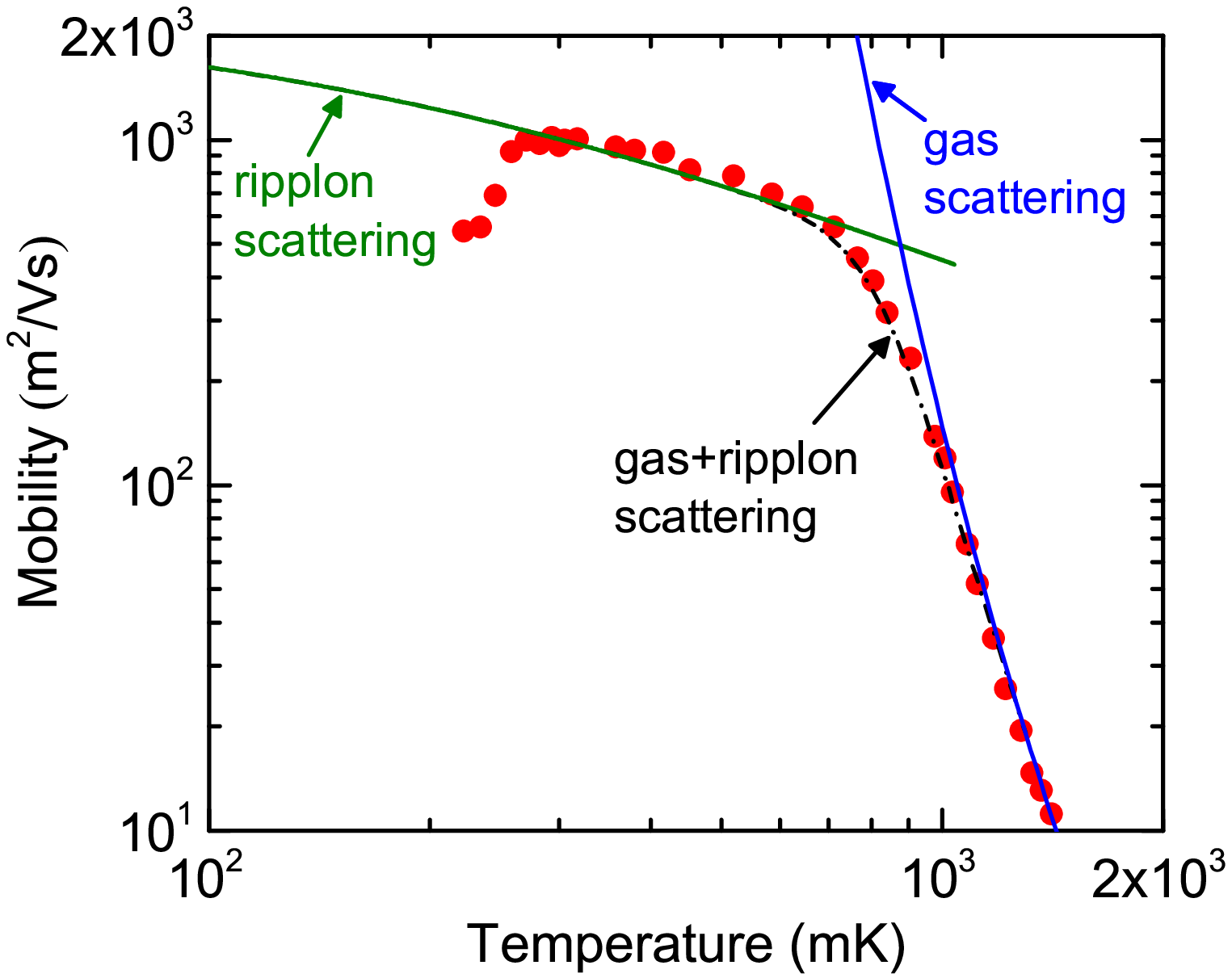}
\end{center}
\caption{\label{mobility_pure4He} Mobility of electrons on the surface of pure liquid $^4$He measured at an electron density of $n_e=$ 1.29$\times$10$^{12}$ m$^{-2}$ and a pressing field of $E_{\bot}=$ 2.09$\times$10$^{4}$ V/m.
The blue line is the theoretical mobility limited by $^4$He gas scattering \cite{Saitoh77}.  
The green line is the ripplon-limited mobility with the many-electron effect [Eq. (3.70) in Ref. \cite{BookMonarkha}].
The dash-dotted line is the theoretical mobility including both the gas scattering and ripplon scattering. 
}
\end{figure}

%\clearpage

%********************************************************************

\section{Theoretical Calculation of Mobility}

Here we describe in detail the theoretical mobility of the WC at a nonzero frequency in the viscous and ballistic regimes presented in the main text.
To calculate the mobility, we consider the motion of a WC in a spatially uniform ac electric field ${\bf{E}}_\parallel  e^{ - i\omega t}$ by noting that the transport properties of the WC dressed with a dimple lattice (DL) are significantly modified from those of bare electrons by the reaction of the helium surface. 
The ac electric field induces uniform displacement of the electrons ${\bf u}  e^{ - i\omega t}$ from their lattice sites, and the displaced electrons experience a reactive force ${\bf{F}}_{\rm D} e^{ - i\omega t}$ from the helium surface.
In this situation, the equation of motion of an electron is described as 
\begin{equation} 
{\bf{F}}_{\rm D} +i m_e\omega \nu_r {\bf u} -e{\bf{E}_\parallel} = - m_e \omega ^2 {\bf u} ,
\label{eq_motion}
\end{equation} 
where direct scattering by thermal ripplons with a rate of $\nu _{r}$ is included.
($e$ is the elementary charge and $m_e$ is the electron mass.)
For a small displacement, the reactive force is linear in the velocity $\dot {\bf u}$ and therefore generally expressed in the form of
\begin{equation} 
{\bf{F}}_{\rm D} = m_e\omega [w(\omega)+i\nu (\omega)] \bf{u} .
\label{F_D}
\end{equation} 
Using the dimensionless response function defined as \cite{Monarkha1997,Monarkha2012,BookMonarkha} 
\begin{equation} 
 Z(\omega)= \left[1+\frac{w(\omega)}{\omega}\right]+i\frac{\nu (\omega)}{\omega}, 
\label{def_Z}
\end{equation} 
the equation of motion is rewritten as
\begin{equation} 
-e{\bf{E}_\parallel} = -m_e \omega ^2 \left[ \mathop{\rm Re}Z(\omega )+ i \left( \mathop{\rm Im}Z(\omega ) + \frac{\nu _{r}}{\omega} \right) \right] {\bf u} .
\label{eq_motion_2}
\end{equation} 
This equation indicates that the real part ${\mathop{\rm Re}\nolimits} Z(\omega)$ represents the change in the inertia of an electron and the imaginary part $\omega {\mathop{\rm Im}\nolimits} Z(\omega)$ describes the momentum relaxation of the electron due to the coupling to the DL.
From Eq. (\ref{eq_motion_2}), the mobility of a WC is expressed as \cite{BookMonarkha,Monarkha1997,Monarkha2012} 
\begin{equation} 
\mu  = \dfrac{e}{m_e}\dfrac{{\omega {\mathop{\rm Im}\nolimits} Z + \nu _{r}}}{{\left( {\omega {\mathop{\rm Re}\nolimits} Z } \right)^2  + \left( {\omega {\mathop{\rm Im}\nolimits} Z + \nu _{r} } \right)^2 }}.
\label{Eq:mobility}
\end{equation} 
This equation suggests that the problem of calculating $\mu$ reduces to evaluating $Z(\omega)$.
Note that ${\mathop{\rm Re}\nolimits} Z(\omega)=0$ at zero frequency, and thus $\mu = \frac{e}{m_e} \frac{1}{{\omega {\mathop{\rm Im}\nolimits} Z + \nu _{r}}}$.

The reactive force from the helium surface arises from the coupling of the electrons to the surface and the dynamics of the surface:  
\begin{equation} 
{\bf{F}}_{\rm D} = -i \sum\limits_{\bf{g}} {\bf g} {\tilde U_g \xi _{\bf g}} e^{{i{\bf g} \cdot {{\bf u} (t)}}} .
\label{F_D_2}
\end{equation} 
Here $\xi _{\bf{g}}$ is the Fourier component of the surface profile $\xi ({\bf r})$ at the reciprocal lattice vector ${\bf{g}}$, $g = \left| \bf{g} \right|$, and $\tilde U_g  = U_g \exp ( - W_g )$ is the electron-ripplon coupling strength, where $U_g=e(E_{\bot}+E_g)$ and $\exp (-W_g)$ is the Debye$-$Waller factor.
($E_{\bot}$ is the pressing field and $E_g$ is the polarization field defined in Refs. \cite{Monarkha1976,BookMonarkha}.)
The Debye$-$Waller factor $\exp (-W_g)$ can be evaluated in the manner described in Refs. \cite{Monarkha1983,BookMonarkha}.

The dynamics of the free surface (i.e., the dynamics of $\xi _{\bf{g}}$) is different in the viscous and ballistic regimes.
Below, we describe the theoretical mobility in both regimes.
We also give details of the direct scattering by thermally excited ripplons.

\subsection{Mobility in Viscous Regime}
In this regime, the dynamics of the surface is determined by the Navier$-$Stokes equation for an incompressible fluid ($\nabla  \cdot {\bf{v}} = 0$) with viscosity $\eta$:
\begin{equation} 
\rho \frac{\partial {\bf{v}}}{{\partial t}} + \nabla p = \eta \Delta {\bf{v}} ,
\label{Navier-Stokes}
\end{equation} 
($\rho$, ${\bf{v}}$, and $p$ are the density, the velocity field, and the pressure of the fluid, respectively) with the following boundary conditions at the free surface associated with the stress tensor $\sigma _{ij}$:
\begin{eqnarray} 
\sigma _{xz}&=&\eta \left( {\frac{{\partial v_x }}{{\partial z}} + \frac{{\partial v_z }}{{\partial x}}} \right) = 0, \nonumber \\
\sigma _{yz}&=&\eta \left( {\frac{{\partial v_y }}{{\partial z}} + \frac{{\partial v_z }}{{\partial y}}} \right) = 0, 
\label{Navier-Stokes_BC} \\
\sigma _{zz}&=&-p +2\eta \frac{{\partial v_z }}{{\partial z}}=\alpha \Delta \xi - P_{e} , \nonumber 
\end{eqnarray} 
where $\alpha$ is the surface tension and $P_{e}$ is the pressure from the electrons.
(The $z$-axis is taken normal to the surface.)
For $P_{e}$, we take the Fourier component and use the linear approximation 
\begin{eqnarray} 
P_{e,\bf g} &=& n_e\tilde U_g e^{-i{{\bf{g}}\cdot {\bf{u}}(t)}} \nonumber \\
&\simeq & n_e \tilde U_g \left[1-i{{\bf{g}}\cdot {\bf{u}}(t)} \right] \equiv P^{(0)}_{{e,\bf g}}+P^{(1)}_{{e,\bf g}}.
\end{eqnarray} 
Solving Eq. (\ref{Navier-Stokes}) with the boundary conditions of Eq. (\ref{Navier-Stokes_BC}), we obtain $\xi$ for $P^{(1)}_{{e,\bf g}}$ as
\begin{equation} 
\xi _{\bf{g}}^{(1)} = -\frac{g P^{(1)}_{{e,\bf g}}}{\rho \Delta _g} ,
\label{xi}
\end{equation} 
where 
\begin{eqnarray} 
\Delta _g &=&  \omega _{r,g}^2 - \omega^2 -\delta _g ^2 -2i\omega \gamma _g, \nonumber \\
\gamma _g &=& \frac{\eta}{\rho}g^2\phi (\kappa) ,   \nonumber \\
\delta _g &=& \omega ^2 \chi (\kappa) , \nonumber \\
\phi (\kappa) &=& 2-\frac{\sqrt 2 }{\kappa} \left[ {\sqrt {1 + \kappa ^2 }  - 1} \right]^{1/2} , \nonumber \\
\chi (\kappa) &=& \frac{4}{\kappa ^2} \left\{ \frac{1}{\sqrt 2 } \left[ {\sqrt {1 + \kappa ^2 }  +1} \right]^{1/2} -1 \right\} , \nonumber 
\end{eqnarray} 
$\kappa = \omega \rho /(\eta g^2) $, and $\omega _{r, g}=\sqrt{\alpha / \rho } g^{3/2}$ is the ripplon frequency with wave number $g$.
(Note that the Fourier components of $\xi ({\bf r})$ in the static state $\xi _g^{(0)} = - \frac{g n_e \tilde U_g}{\rho \omega _{r, g}^2}$ are obtained from $P^{(0)}_{{e,\bf g}}$.)
Substituting Eq. (\ref{xi}) into Eq. (\ref{F_D_2}) and using Eqs. (\ref{F_D}) and (\ref{def_Z}), we obtain \cite{Monarkha1997,Monarkha2012,BookMonarkha}
\begin{equation} 
Z(\omega) = 1+ \frac{\rho}{2m_e n_e}\sum\limits_{\bf{g}} {g\left| {\xi _g^0 } \right|^2 \left( {\frac{{\omega _{r, g}}}{\omega }} \right)^2 \frac{{\omega ^2  + \delta _g ^2  + i2\omega \gamma _g }}{{\Delta _g }}
} ,
\label{eq_mu_v}
\end{equation} 
where $n_e$ is the density of electrons. 
To calculate the theoretical mobility in the viscous regime described in the main text, we first evaluate $Z(\omega)$ [Eq. (\ref{eq_mu_v})] using the interpolated values of $\rho$, $\alpha$, and $\eta$ of the mixture presented in Sec. \ref{interpolation} of this Supplementary Information, and then evaluate $\mu$ using Eq. (\ref{Eq:mobility}).
Note that at the zero-frequency limit,  $\omega {\mathop{\rm Re}\nolimits} Z(\omega) =0$ and $\omega {\mathop{\rm Im}\nolimits} Z= \nu _d= \frac{\eta }{{m_en_e }}\sum\limits_{\bf{g}} {g^3 } \left| {\xi _{g}^0 } \right|^2 $, indicating $\mu$ to be proportional to $\eta^{-1}$.

\subsection{Mobility in Ballistic Regime}
In this regime, the mobility of the WC is caused by the reflection of QPs.
The reflection of QPs generates a drag force on the moving DL, and the drag force is transferred to the WC. 
At a nonzero frequency, we must also include the inertial term in the calculation of the mobility, which can be performed by using the response function $Z (\omega )$.
The response function can be obtained from the dynamics of the free surface by noting that microscopic information of the reflection of QPs is incorporated in the damping of a capillary wave.
We consider the dynamics of the free surface under the damping of a capillary wave, which is described by 
\begin{equation} 
\ddot \xi _{\bf{g}} +2\gamma_g \dot \xi _{\bf{g}} + \omega _{r,g}^2 \xi _{\bf{g}} = -\frac{\tilde U_g n_e g}{\rho} e^{-i{\bf{g}} \cdot {\bf{u}}(t)} ,
\label{eq_motion_surface}
\end{equation} 
where $\gamma_{g}$ is the damping rate of a capillary wave with wave number $g$.
Solving Eq. (\ref{eq_motion_surface}) for a small displacement $\bf{u}$ and using Eqs. (\ref{F_D}), (\ref{def_Z}), and (\ref{F_D_2}), we obtain \cite{Monarkha2012,BookMonarkha}
\begin{equation} 
 Z (\omega ) = 1+ \sum\limits_{\bf{g}} \dfrac{{n_e \tilde U_{g }^2 g}}{{m_e \rho \omega _{r,g }^2 }} g_x^2 \dfrac{{\left( {\omega _{r,g }^2  - \omega ^2 } - 4\gamma _{g }^2 \right) +2i \gamma _{g} {\omega _{r,g }^2} / \omega}}{{\left( {\omega _{r,g }^2  - \omega ^2 } \right)^2  + 4\gamma _{g }^2 \omega ^2 }} ,
\label{Eq:Z_QP}
\end{equation} 
where $g_x$ is the component of $\bf{g}$ parallel to ${\bf v} = \dot {\bf u}$.
Note that $\gamma_g$ includes information on the microscopic nature of the reflection of a QP.

In the case of specular reflection,  the damping rate $\gamma_g^{(s)}$ caused by the reflection of QPs is given by \cite{Monarkha1997,Monarkha2005,BookMonarkha}
\begin{equation} 
\gamma_g^{(s)} =  - \frac{g}{{8\pi ^2 \hbar ^3 \rho }}\int\limits_0^\infty  {p^4 \frac{{df(\varepsilon )}}{{dp}}dp} ,
\label{gamma_specular} 
\end{equation}
where $\hbar$ is Planck's constant and $f(\varepsilon )$ is the Fermi distribution function of $^3$He QPs with energy $\varepsilon = p^2/(2m_{\rm 3D}^*)$ ($p$ is the momentum and $m_{\rm 3D}^*$ is the effective mass of a $^3$He QP). 
Note that at $T  \ll T_F$, the mobility is independent of temperature $T$ [$\gamma_g^{(s)} = g p_F^4/(8\pi ^2 \hbar ^3 \rho)$] ($T_F$ is the Fermi temperature and $p_F$ is the Fermi momentum).
From Eq. (\ref{gamma_specular}), we obtain $Z (\omega )$ using Eq. (\ref{Eq:Z_QP}), with which we calculate the mobility using Eq. (\ref{Eq:mobility}).
Note that at $\omega \to 0$,  $\omega {\mathop{\rm Re}} Z \to 0$ and 
\begin{equation} 
\omega {\mathop{\rm Im}} Z \to \frac{{\kappa _s (T)}}{{m_en_e }}\sum\limits_{\bf{g}} {g_x^2 } \left| {\xi _g^0 } \right|^2 ,
\label{Eq:mu_specular}
\end{equation} 
reproducing the zero-frequency mobility given in Ref. \cite{Monarkha1997}, where  
\begin{equation} 
\kappa_s (T) =  - \frac{1}{{4\pi ^2 \hbar ^3 }}\int_0^\infty  {p^4 } \frac{{df(\varepsilon )}}{{dp}}dp .
\label{Eq:kappa_s}
\end{equation}

In the case of the reflection associated with the complete accommodation of a QP into the surface layer, the damping rate is given by \cite{Monarkha2006}
\begin{equation} 
\gamma _g^{(a)} (T) = \frac{1}{4} \gamma _g^{(s)} (T).
\label{Eq:mu_diffusive}
\end{equation} 
The mobility in the accommodation process presented in the main text is calculated using Eqs. (\ref{Eq:Z_QP}) and (\ref{Eq:mu_diffusive}).
Note that $\gamma _g^{(a)} (T)$ and $\gamma_g^{(s)} (T)$ have the same function of $T$ but are different by a factor of four.
Thus, at zero frequency, the mobility limited by the accommodation process is four times higher than that in the case of specular reflection.
At a nonzero frequency, the term $\omega {\mathop{\rm Re}} Z (\ne 0)$ contributes to the mobility, resulting in the difference in mobilities between the specular and accommodation processes being less than a factor of four.

In the practical situation at the surface of the mixture, not all but a part of the $^3$He QPs with ratio $r_a$ are reflected by the accommodation process.
In this partial accommodation process, the damping rate is described as
\begin{equation} 
\gamma_g (T)= \left( {1 - \frac{3}{4} r_a } \right) \gamma_g^{(s)} (T) , 
\label{Eq:damping_diffusive}
\end{equation}
where $r_a=0$ corresponds to specular reflection and $r_a=1$ to complete accommodation.
In the main text, the experimental data are fitted to this partial accommodation model with the fitting parameter $r_a$.

\subsection{Contribution from Direct Scattering by Thermal Ripplons}
\label{direct_ripplon_Scattering}
In the theoretical mobility presented in the main text, we also include the contribution of the direct scattering by thermally excited ripplons.
We evaluate the rate of collision with the ripplons $\nu _{r}$ using the Born approximation.
In this approximation, $\nu _{r}$ is described as a simple form including the dynamical structure factor (DSF).
For the DSF of the WC, we use the high-temperature approximation [Eq. (8.2) in Ref. \cite{BookMonarkha}], where the form of the DSF is the same as that of a non-degenerate electron gas with the temperature replaced by the kinetic energy of electrons in the WC.

\bibliography{suppl.bib}